\documentclass[preprint,amsmath,amssymb,nofootinbib]{revtex4}
\usepackage[utf8]{inputenc}
\usepackage{rotating}
\usepackage{hyperref}
\usepackage{array}
\usepackage{subfig}
\usepackage{float}
\usepackage{graphicx}
\usepackage{slashed}
\usepackage{color}
\definecolor{orange}{rgb}{1,0.5,0}
\usepackage[T1]{fontenc}

\newcommand{\tvect}[2]{%
  \ensuremath{\Bigl(\negthinspace\begin{smallmatrix}#1\\#2\end{smallmatrix}\Bigr)}}

\begin{document}

\title{Majorana neutrinos with effective interactions in $B$ decays.}

\author{Luc\'{\i}a Duarte}
\email{lduarte@fing.edu.uy}
 \affiliation{Instituto de F\'{\i}sica, Facultad de Ingenier\'{\i}a,
 Universidad de la Rep\'ublica \\ Julio Herrera y Reissig 565,(11300) 
Montevideo, Uruguay}
 \affiliation{Instituto de F\'{\i}sica, Facultad de Ciencias,
 Universidad de la Rep\'ublica \\ Igu\'a 4225,(11400) 
Montevideo, Uruguay.}

\author{Javier Peressutti}
\email{javier.peressutti@mdp.edu.ar}
\affiliation{Instituto de F\'{\i}sica de Mar del Plata (IFIMAR)\\
CONICET, UNMDP\\ Departamento de F\'{\i}sica,
Universidad Nacional de Mar del Plata \\
Funes 3350, (7600) Mar del Plata, Argentina}

\author{Ismael Romero}
\affiliation{Instituto de F\'{\i}sica de Mar del Plata (IFIMAR)\\
CONICET, UNMDP\\ Departamento de F\'{\i}sica,
Universidad Nacional de Mar del Plata \\
Funes 3350, (7600) Mar del Plata, Argentina}

\author{Oscar A. Sampayo}
\email{sampayo@mdp.edu.ar}
\affiliation{Instituto de F\'{\i}sica de Mar del Plata (IFIMAR)\\ CONICET, UNMDP\\ Departamento de F\'{\i}sica,
Universidad Nacional de Mar del Plata \\
Funes 3350, (7600) Mar del Plata, Argentina}

\begin{abstract}
We investigate the possible contribution of Majorana neutrinos to $B$ meson decays in an effective interaction formalism, in the mass range $0.5$ GeV $<m_N<5 $ GeV. We study the decay of the $B^-$ meson via $B^- \to  \mu^- \mu^- \pi^+$ at LHCb, which is a signal for leptonic number violation and the presence of Majorana neutrinos, and put bounds on different new physics contributions, characterized by their Dirac-Lorentz structure. We also study the bounds imposed by the radiative $B$ decay ($B^- \rightarrow \mu^- \nu \gamma$) results from Belle. The obtained bounds are more restrictive than previous values found for dimension 6 four-fermion contact vectorial and scalar Majorana neutrino interactions in the context of the Left-Right symmetric model for higher Majorana masses at the LHC, showing that the direct calculation of the effective $N$ interactions contribution to different processes can help to put more stringent bounds to different UV-complete models parameterized by an effective Lagrangian. 
 \end{abstract}

\maketitle

\section{Introduction.}

The search for particles beyond the standard model (SM) content has been extensive in the past few years, among them sterile Majorana neutrinos $N$, which appear as a natural consequence in several SM extensions. The discovery of neutrino oscillations suggests that the standard neutrinos are massive particles. One of the possible ways to generate their mass is the seesaw mechanism \cite{Minkowski:1977sc, Mohapatra:1979ia, Yanagida:1980xy, GellMann:1980vs, Schechter:1980gr}, which introduces at least one right handed singlet and produces Majorana neutrinos. In this way one obtains masses for the standard neutrinos $m_\nu \sim Y/M_N$ of order $10^{-2}~ eV$ compatible with current oscillation data, assuming sufficiently heavy Majorana masses ($M_N \sim 10^{15}$ GeV) and convenient Yukawa couplings of order $Y\sim 1$. On the other hand, for smaller Yukawa couplings of order $Y\sim 10^{-8}-10^{-6}$, sterile neutrinos with masses around $M_{N}\sim (1-1000) $ GeV could exist. However, in the simplest Type-I seesaw scenarios, a major drawback is that the left-right mixing parameters $U_{lN}\, (l=e,\,\mu\,,\tau)$ need to be negligibly small $U_{lN}^2 \sim m_{\nu}/M_N \sim 10^{-14}-10^{-10}$ in order to account for light $\nu$ masses \cite{Cai:2017mow, Atre:2009rg}. The mixings $U_{lN}$ weight the coupling of the heavy $N$ with the SM particles, in particular with charged leptons through the $V-A$ interaction
\begin{equation}\label{eq:mix}
  \mathcal{L}^W_{V-A} = -\frac{g}{\sqrt{2}} U_{lN} \overline{N^c}\gamma^{\mu}P_L l W^+  +  {\rm h.c.},
\end{equation}
so this leads to the decoupling of the Majorana neutrinos. However, the observation of any lepton number violating (LNV) process would point to the Majorana nature of the exchanged fermion. Recent approaches consider a toy-like model in which the SM is extended by incorporating a massive Majorana sterile fermion, assumed to have non-negligible mixings with the active states, without making any hypothesis on the neutrino mass generation mechanism \cite{Abada:2017jjx, Pascoli:2018heg}. Such a minimal SM extension leads to contributions to LNV observables which are already close, or even in conflict, with current data from meson and tau decays, for Majorana masses $M_N$ below $10$ GeV (see \cite{Abada:2017jjx, Abada:2018nio} and the references therein). So, also from the experimental point of view, the simple SM extensions which attribute LNV only to the mixing between heavy Majorana states and the active neutrinos are facing increasingly stringent constraints.

As suggested in \cite{delAguila:2008ir}, the detection of Majorana neutrinos ($N$) would be a signal of physics beyond the minimal seesaw mechanism, and its interactions could be better described in a model independent approach based on an effective theory. One can think of an alternative treatment and consider the Majorana neutrino interactions as originating in new physics described by an unknown underlying renormalizable theory valid at a higher energy (UV) scale and parametrized at low energies by a model independent effective Lagrangian. In this approach, we consider that the sterile $N$ interacts with the SM particles by higher dimension effective operators, taking these interactions to be dominant in comparison with the mixing with light neutrinos through the Yukawa couplings, which we neglect \cite{Peressutti:2011kx,Peressutti:2014lka,Duarte:2014zea,Duarte:2015iba,Duarte:2016miz,Duarte:2016caz,Duarte:2018xst,Duarte:2018kiv}. We depart from the usual viewpoint in which the mixing with the standard neutrinos is assumed to govern the $N$ production and decay mechanisms. Here, for simplicity, we consider a scenario with only one Majorana neutrino $N$ and negligible mixing with the $\nu_{L}$. 

The different operators in the effective Lagrangian, with distinct Dirac-Lorentz structure, parameterize a wide variety of UV-complete new physics models, like extended scalar and gauge sectors as the Left-Right symmetric model, vector and scalar leptoquarks, etc. Thus, discerning between the possible contributions given by them to specific processes gives us a hint on what kind of new physics at a higher energy regime could be responsible for the observed interactions.

Observable effects of the existence of sterile Majorana neutrinos such as lepton number violation have been sought thoroughly in hadron colliders like the LHC, $e^+e^-$ and $ep$ colliders, low energy high precision experiments as neutrinoless double beta decay searches ($0\nu\beta\beta$) among others (for comprehensive reviews see \cite{Cai:2017mow, Deppisch:2015qwa} and references therein). In particular, heavy flavor meson decays could be the place where for the first time the Majorana neutrino effects were observed or, in the absence of a discovery, this fact can be used to set limits for its coupling to SM particles. $N$-mediated lepton number violation in rare $B$ meson decays has been studied, for example, in \cite{Abada:2017jjx, Asaka:2016rwd, Cvetic:2016fbv, Cvetic:2015naa, Wang:2014lda, Cvetic:2010rw, Zhang:2010um, Helo:2010cw, Atre:2009rg, Ali:2001gsa}, and the references therein. Concerning the resonant production of Majorana neutrinos in semileptonic pseudoscalar meson three-body decays, the recently measured branching ratio $Br(B^- \to  \mu^- \mu^- \pi^+)< 4\times 10^{-9}$ for intermediate neutrinos with lifetimes $\tau_N$ shorter than $1$ ps at the LHCb experiment \cite{Aaij:2014aba} gives the currently more stringent bounds on the mixing parameter $|U_{\mu N}|^{2}$ in the case of the minimal SM extension by one Majorana neutrino (e.g. \cite{Abada:2017jjx, Yuan:2017uyq}) for Majorana masses in the range $2.5\lesssim M_N \lesssim 5 $ GeV. 

In this paper we aim to exploit the recent $B$-decay data to constrain the possible values of the couplings that weight the contribution of different effective operators to the Majorana-mediated same sign dilepton $B$-decay $B^{-} \to \mu^{-} \mu^{-} \pi^{+}$ and the radiative leptonic muon-mode $B^{-} \to \mu^{-} \nu \gamma$. The LHCb collaboration has presented model independent upper limits on the branching ratio of the first process \cite{Aaij:2014aba, Ossowska:2018ybk}, and the Belle collaboration has set new limits on the integrated differential width of the $B^{-} \to \mu^{-} \nu \gamma$ decay \cite{Gelb:2018end}. The obtained bounds (for $0.5 \lesssim m_N \lesssim 5 $ GeV) are more restrictive than previous values obtained for dimension 6 four-fermion contact vectorial and scalar Majorana neutrino interactions in the context of the Left-Right symmetric model for higher Majorana masses \cite{Ruiz:2017nip}, and constrain the perspectives of discovery of Majorana neutrinos with effective interactions with GeV-scale masses by direct production in colliders and meson decays \cite{Duarte:2016caz, Yue:2017mmi, Yue:2018hci}. 

The paper is organized as follows. In Sect. \ref{sec:eff} we introduce the effective Lagrangian formalism. In Sect. \ref{sec:NmedBdec} we present the analytical results for the $B^- \rightarrow \pi^+ \mu^- \mu^-$ branching ratio and the $B^- \rightarrow \mu^- \nu \gamma$ decay in this formalism. In Sect. \ref{sec:results} we discuss our numerical results for the obtained bounds, and in Sect.\ref{sec:final} we make our final comments. The details of the calculations are presented in Appendices \ref{sec:bdec_muN}, \ref{sec:bdec_pion} and \ref{sec:ap_bdec_rad}.

\subsection{Majorana neutrino effective interactions.}\label{sec:eff}

An appropriate way to include the Majorana neutrino into the theory is to extend the SM Lagrangian. In this work we consider an effective Lagrangian in which we include only \emph{one} relatively light right handed Majorana neutrino $N$ as an observable degree of freedom. The new physics effects are parameterized by a set of effective operators $\mathcal{O}_\mathcal{J}$ constructed with the SM and the Majorana neutrino fields and satisfying the $SU(2)_L \otimes U(1)_Y$ gauge symmetry \cite{Wudka:1999ax}. 

The effect of these operators is suppressed by inverse powers of the new physics scale $\Lambda$. The total Lagrangian is organized as follows:

\begin{eqnarray}\label{eq:Lagrangian}
\mathcal{L}=\mathcal{L}_{SM}+\sum_{n=5}^{\infty}\frac1{\Lambda^{n-4}}\sum_{\mathcal{J}} \alpha_{\mathcal{J}} \mathcal{O}_{\mathcal{J}}^{(n)}
\end{eqnarray}
where $n$ is the mass dimension of the operator $\mathcal{O}_{\mathcal{J}}^{(n)}$.

Note that we do not include the Type-I seesaw Lagrangian terms giving the Majorana and Yukawa terms for the sterile neutrinos. The dominating effects come from the lower dimension operators that can be generated at tree level in the unknown underlying renormalizable theory.

The dimension 5 operators were studied in detail in \cite{Aparici:2009fh}. These include the Weinberg operator $\mathcal{O}_{W}\sim (\bar{L}\tilde{\phi})(\phi^{\dagger}L^{c})$ \cite{Weinberg:1979sa} which contributes to the light neutrino masses, $\mathcal{O}_{N\phi}\sim (\bar{N}N^{c})(\phi^{\dagger} \phi)$ which gives Majorana masses and couplings of the heavy neutrinos to the Higgs (its LHC phenomenology has been studied in \cite{Caputo:2017pit, Graesser:2007yj}), and the operator $\mathcal{O}^{(5)}_{NB}\sim (\bar{N}\sigma_{\mu \nu}N^{c}) B^{\mu \nu}$ inducing magnetic moments for the heavy neutrinos, which is identically zero if we include just one sterile neutrino $N$ in the theory.

In the following, as the dimension 5 operators do not contribute to the studied processes -discarding the heavy-light neutrino mixings- we will only consider the contributions of the dimension 6 operators, following the treatment presented in \cite{delAguila:2008ir}. We start with a rather general effective Lagrangian density for the interaction of right-handed Majorana neutrinos $N$ including dimension 6 operators.

The first operators subset includes those with scalar and vector bosons (SVB), 
\begin{eqnarray} \label{eq:Ope-SVB}
\mathcal{O}^{(i)}_{LN\phi}=(\phi^{\dag}\phi)(\bar L_i N \tilde{\phi}),
\;\; \mathcal{O}_{NN\phi}=i(\phi^{\dag}D_{\mu}\phi)(\bar N \gamma^{\mu} N), \;\; \mathcal{O}^{(i)}_{Nl\phi}=i(\phi^T \epsilon D_{\mu}\phi)(\bar N \gamma^{\mu} l_i)
\end{eqnarray}
and a second subset includes the baryon-number conserving four-fermion contact terms (4-f):  
\begin{eqnarray} \label{eq:Ope-4-f} 
\mathcal{O}^{(i)}_{duNl}&=&(\bar N \gamma_{\mu} l_i) (\bar d_i \gamma^{\mu} u_i) , 
\;\; \mathcal{O}^{(i)}_{fNN}=(\bar N \gamma_{\mu}N) (\bar f_i \gamma^{\mu}f_i), 
\;\; \mathcal{O}^{(i)}_{LNLl}=(\bar L_i N)\epsilon (\bar L_i l_i),
\nonumber \\
\mathcal{O}^{(i)}_{LNQd}&=&(\bar L_i N) \epsilon (\bar Q_i d_i), \;\; \mathcal{O}^{(i)}_{QuNL}=(\bar Q_i u_i)(\bar N L_i) , \;\; \mathcal{O}^{(i)}_{QNLd}=(\bar Q_i N)\epsilon (\bar L_i d_i), 
\nonumber \\
\mathcal{O}^{(i)}_{L N}&=&|\bar N L_i|^2 , \;\; \mathcal{O}^{(i)}_{QN}=|\bar Q_i N|^2
\end{eqnarray}
where $l_i$, $u_i$, $d_i$ and $L_i$, $Q_i$ denote, for the family labeled $i$, the right handed $SU(2)$ singlet and the left-handed
$SU(2)$ doublets, respectively. The field $\phi$ is the scalar doublet.
Also $\gamma^{\mu}$ are the Dirac matrices, and $\epsilon=i\sigma^{2}$ is the antisymmetric symbol.

One can also consider operators generated at one-loop (1-loop) order in the underlying full theory, whose coefficients are naturally suppressed by a factor $1/16\pi^2$\cite{delAguila:2008ir, Arzt:1994gp}:
\begin{eqnarray}\label{eq:Ope-1-loop} 
\mathcal{O}^{(i)}_{ N B} = (\bar L_i \sigma^{\mu\nu} N) \tilde \phi B_{\mu\nu} , &&
\mathcal{O}^{(i)}_{ N W } = (\bar L_i \sigma^{\mu\nu} \tau^I N) \tilde \phi W_{\mu\nu}^I
\end{eqnarray}
Here $B_{\mu\nu}$ and $W_{\mu\nu}^I$ represent the $U(1)_{Y}$ and $SU(2)_{L}$ field strengths respectively, and $\sigma^{\mu \nu}$ is the Dirac tensor. 

The effective operators above can be classified by their Dirac-Lorentz structure into \emph{scalar}, \emph{vectorial} and \emph{tensorial}. 
The complete expression for the effective dimension 6 Lagrangian terms \footnote{It must be stressed that the Majorana neutrino $N$ is right-handed, so all the vector terms involve a ($V+A$) interaction.} can be found in Appendix A in \cite{Duarte:2016miz}. 

In this paper we will consider the $B$ decays $B^-\rightarrow \mu^-\mu^-\pi^+$ in Sect. \ref{sec:Nmupi} and $B^- \rightarrow \mu^- \nu \gamma$ in Sect. \ref{sec:bdec_radiative}, mediated by an on-shell Majorana neutrino $N$. We can thus take into account the following effective Lagrangian terms involved in the $B^- \rightarrow \mu^- N$  and $N\rightarrow \mu^-\pi^+$ processes (from eqs. \eqref{eq:Ope-SVB} and  \eqref{eq:Ope-4-f}):  
\begin{equation}\label{eq:efflag}
  \mathcal{L} = \mathcal{L}_{SM} + \frac{1}{\Lambda^2}   \big(   \alpha^{(i)}_{Nl\phi}\, \mathcal{O}_{Nl\phi} 	+
  \alpha^{(i)}_{QuNL} \, \mathcal{O}_{QuNL}+
\alpha^{(i)}_{duNl}\, \mathcal{O}_{duNl} +\alpha^{(i)}_{LNQd}\, \mathcal{O}_{LNQd}	+ \alpha^{(i)}_{QNLd}\, \mathcal{O}_{QNLd} \big).
\end{equation}
The couplings $\alpha^{(i)}_{\mathcal O}$ are associated to specific operators:
\begin{eqnarray}
\alpha^{(i)}_W&=&\alpha^{(i)}_{N l\phi},\; \; 
\alpha^{(i)}_{V_0}=\alpha^{(i)}_{duNl}, \nonumber \\
\alpha^{(i)}_{S_1}&=&\alpha^{(i)}_{QuNL},\; \alpha^{(i)}_{S_2}=\alpha^{(i)}_{LNQd},\;\;
\alpha^{(i)}_{S_3}=\alpha^{(i)}_{QNLd}.
\end{eqnarray}
After spontaneous electroweak symmetry breaking, taking the scalar doublet as $\phi=\tvect{0}{\frac{v+h}{\sqrt{2}}}$, with $h$ being the Higgs field and $v$ its vacuum expectation value, we can write the Lagrangian \eqref{eq:efflag} terms involved in our calculation (and its charge conjugate), as
\begin{eqnarray}\label{eq:lag}
  \mathcal{L}&& = \mathcal{L}_{SM} + \frac{1}{\Lambda^2} \Big\{ -\alpha^{(i)}_W \frac{ ~v ~m_W}{\sqrt{2}}\, \overline{l_i} \gamma^{\nu} P_R N  \, W^{-}_{\mu}    
		+  \alpha^{(i)}_{V_0} \, \overline{u'_i} \gamma^{\nu} P_R d'_i \,  \, \overline{l_i} \gamma_{\nu} P_R N   \nonumber \\
		&& +  \alpha^{(i)}_{S_1}\, \, \overline{u'_i} P_L d'_i \, \, \overline{l_i} P_R N -  \alpha^{(i)}_{S_2}\, \overline{u'_i} P_R d'_i \, \, \overline{l_i} P_R N  + \alpha^{(i)}_{S_3}\, \overline{u'_i} P_R N \, \,  \overline{l_i} P_R d'_i + \mbox{h.c.} \Big\}.
\end{eqnarray}
Here the quark fields are flavor eigenstates with family $i=1,2,3$. In order to find the contribution of the effective Lagrangian to the $B^{-} \rightarrow \mu^{-} N$ and $N \rightarrow \mu^{-} \pi^{+}$ decays, we must write it in terms of the massive quark fields. 
Thus, we consider that the contribution of the dimension 6 effective operators to the Yukawa Lagrangian are suppressed by the new physics scale with a factor $\frac{1}{\Lambda^2 }$, and neglect them, so that the matrices that diagonalize the quark mass matrices are the same as in the pure SM. 

Writing with a prime symbol the flavor fields, we take the matrices $U_{R}, ~U_{L}, D_{R}$ and $D_{L}$ to diagonalize the SM quark mass matrix in the Yukawa Lagrangian. Thus the left- and right- handed quark flavor fields (subscript $i$) are written in terms of the massive fields (subscript $\beta$) as:
\begin{eqnarray}\label{eq:mass_basis}
&& u^{'}_{(R,L) i} = U_{(R,L)}^{i,\beta} u_{(R,L) \beta}  \;,  \; \; \; \; \; \overline{u^{'}_{(R,L)i}}=\overline{u_{(R,L)\beta}} (U_{(R,L)}^{i, \beta})^{\dagger}    \nonumber \\
&& d^{'}_{(R,L) i} = D_{(R,L)}^{i,\beta} d_{(R,L) \beta} \;,  \; \; \; \; \; \overline{d^{'}_{(R,L)i}}=\overline{d_{(R,L)\beta}} (D_{(R,L)}^{i, \beta})^{\dagger} .
\end{eqnarray}
With this notation, the SM $V_{CKM}$ mixing matrix corresponds to the term $V^{\alpha \beta} = \sum_{i=1}^{3} (U_{L}^{i \alpha})^{\dagger}D_L^{i \beta} $ appearing in the charged SM current $J^{\mu}_{CC}= \overline{u^{\alpha}_L} ~V^{\alpha \beta} \gamma^{\mu}  P_L   d_L^{\beta}$.

For the $N\rightarrow \nu \gamma$ decay, the considered Lagrangian terms come from one-loop level generated tensorial operators: 
\begin{eqnarray}\label{eq:leff_1loop_L}
\mathcal{L}_{eff}^{1-loop}&=& \frac{-i\sqrt{2} v}{\Lambda^2} {(\alpha_{NB}^{(i)}c_W + 
 \alpha_{NW}^{(i)}s_W)  (P^{(A)}_{\mu} ~\bar \nu_{L,i} \sigma^{\mu\nu}N_R~ A_{\nu})}.
\end{eqnarray}
where $-P^{(A)}$ is the 4-momentum of the outgoing photon, $s_W$ and $c_W$ are the sine and cosine of the weak mixing angle, and a sum over the neutrino family index $i$ is understood. The couplings $\alpha^{(i)}_{NB}$ and $\alpha^{(i)}_{NW}$ correspond respectively to the operators in \eqref{eq:Ope-1-loop}.


\section{$N$ mediated $B$ decays.}\label{sec:NmedBdec}

\begin{figure*}[h]
 \centering
  \includegraphics[width=0.8\textwidth]{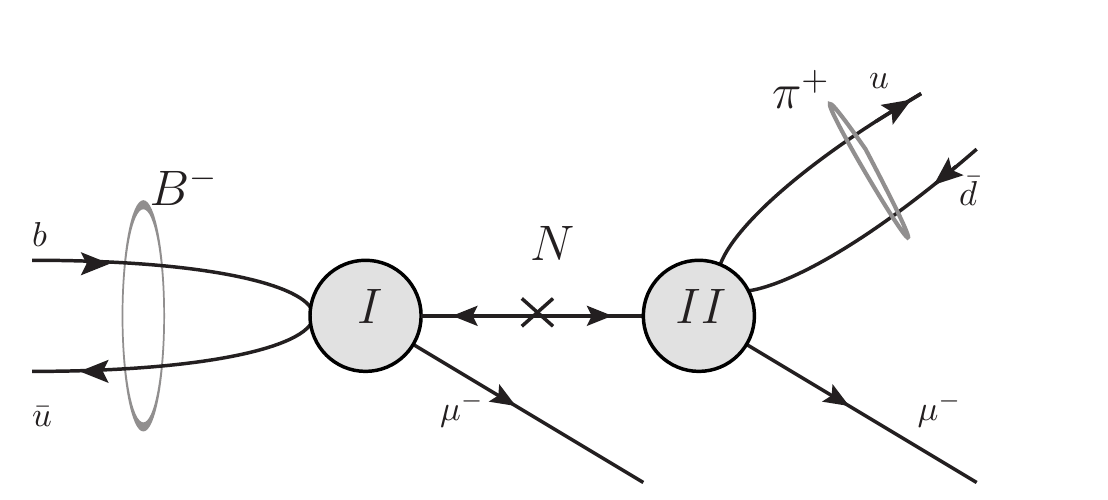}
 \caption{\label{fig:Bmumupi} Schematic representation for the effective contribution to the decay $B^-\rightarrow  \mu^-\mu^- \pi^+$.}
 \end{figure*}

We first consider the lepton number violating $B^{-} \to \mu^{-} \mu^{-} \pi^{+}$ decay shown in Fig. \ref{fig:Bmumupi}. This LNV process is strictly forbidden in the SM and when mediated by a Majorana neutrino $N$ it allows to probe masses up to $m_N=5 $ GeV. Also, the radiative muon-mode $B^{-} \to \mu^{-} \nu \gamma$ shown in Fig. \ref{fig:Bmunugamma} is well suited to probe this mass range, as the $N \to \nu \gamma$ channel dominates the total $N$ decay with for Majorana masses up to $30$ GeV \cite{Duarte:2015iba}.

We calculate the decay of the $B^-$ meson for the two studied processes in two stages. Firstly we obtain the decay of $B^-$ to a muon $\mu^-$ and a Majorana neutrino $N$. Secondly, we calculate the decay of $N \to \mu^- \pi^+$ and the radiative decay $N \rightarrow \nu \gamma$. 

\begin{figure*}
 \centering
  \includegraphics[width=0.8\textwidth]{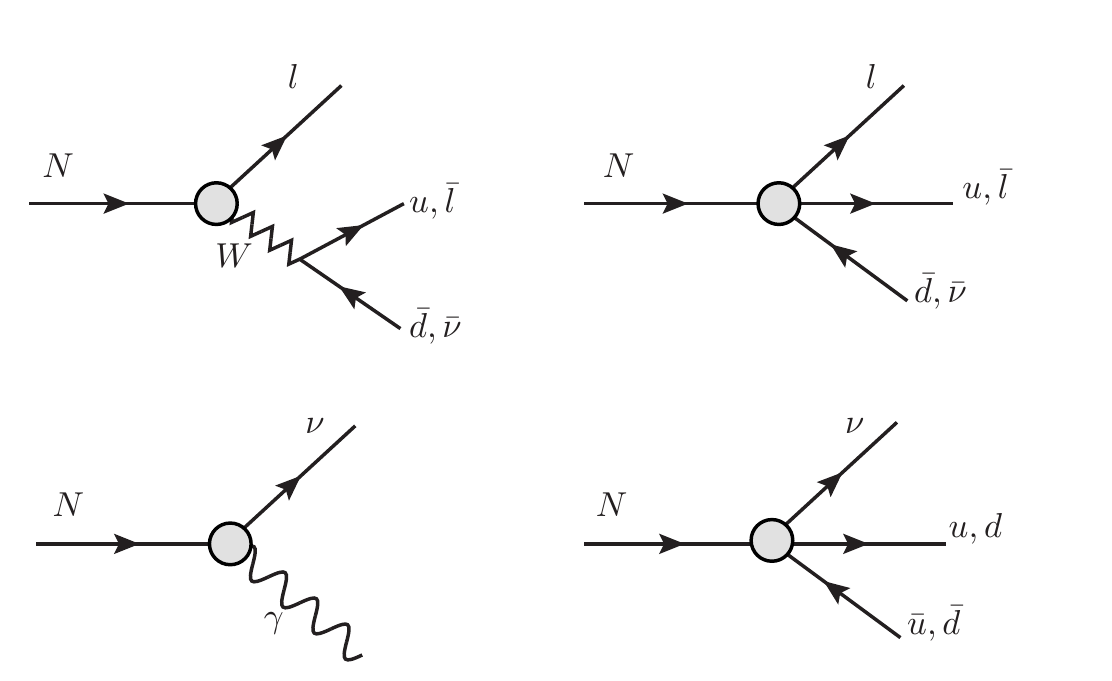}
 \caption{\label{fig:N_decay} Schematic representation for the low mass Majorana neutrino decay channels.}
 \end{figure*}

The decay width of the $B^-$ meson is obtained in both cases in the following way
\begin{equation}
  \Gamma(B^-\to \mu^- \mu^- \pi^+) = \Gamma(B^-\to \mu^-N)\, Br(N\to \mu^- \pi^+), \label{eq:Btopi_decay} 
\end{equation}
\begin{equation}
	\mbox{with}\qquad Br(N\to \mu^- \pi^+) = \Gamma(N\to \mu^- \pi^+)/\Gamma_N   \label{eq:BrNtopi} 
\end{equation}
and
\begin{equation}
  \Gamma(B^-\to \mu^- \nu \gamma) = \Gamma(B^- \to \mu^- N)\, Br(N \to \nu \gamma), \label{eq:Btomunug_decay}
\end{equation}
\begin{equation}
	\mbox{with}\qquad Br(N\to \nu \gamma) = \Gamma(N\to \nu \gamma)/\Gamma_N,    \nonumber
\end{equation}
where $\Gamma_N$ is the total decay width for the Majorana neutrino. This is equivalent to calculating the whole decay process assuming an on-shell intermediate Majorana neutrino. For the $N$ decay width we include all the kinematically allowed channels for a Majorana neutrino of mass in the range $0.5  ~GeV  <m_N<5  $ GeV which are depicted in Fig.\ref{fig:N_decay}. In our calculation we keep all the final-state masses.

\begin{figure*}
\centering
\subfloat[Sets 1, 2, 3]{\label{fig:ancho_set123}\includegraphics[totalheight=6.cm]{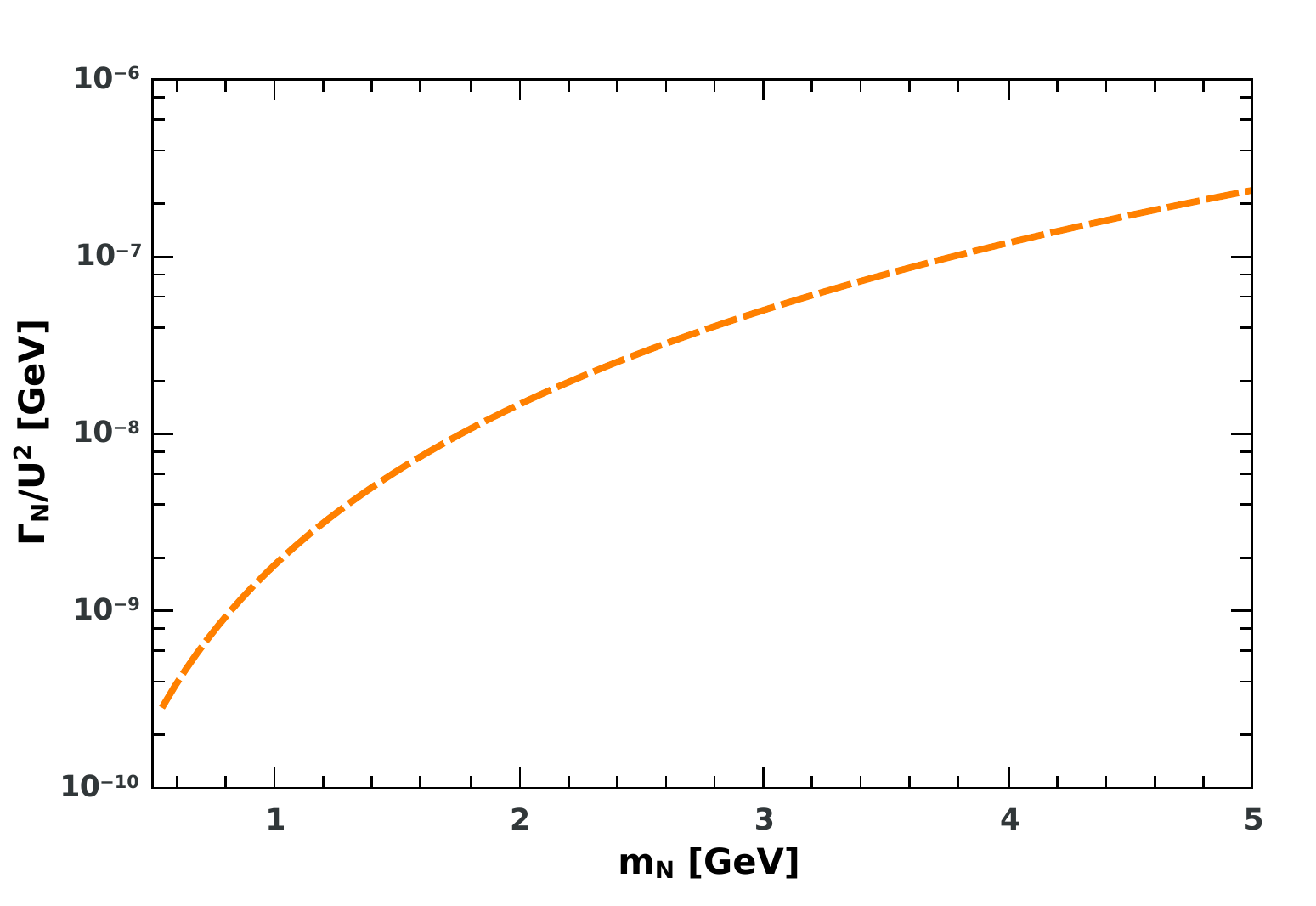}}~
\subfloat[Sets 4, 5, 6]{\label{fig:ancho_set456}\includegraphics[totalheight=6.cm]{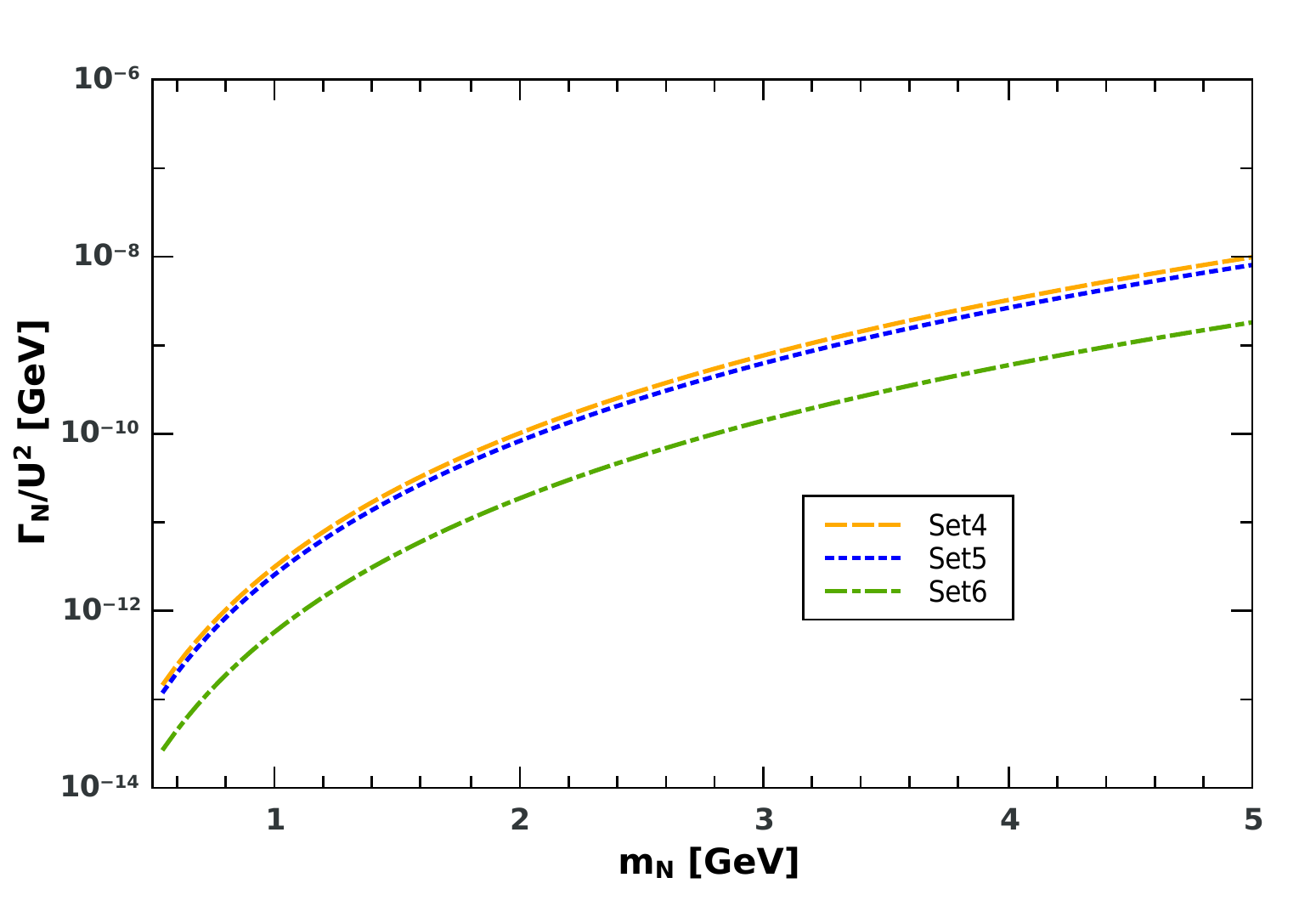}}
\caption{Total $\Gamma_N$ width for different coupling sets in Tab.\ref{tab:alpha-sets}. The curves for sets 1, 2 and 3 overlap.}\label{fig:ancho}
\end{figure*}

The details of the calculation of the total $N$ decay width are described in \cite{Duarte:2016miz}. In Figs.\ref{fig:ancho_set123} and \ref{fig:ancho_set456} we present the results for the total width $\Gamma_N$ for the different sets of effective couplings, as will be described in Sect. \ref{sec:numerical}.

%

\subsection{The $B^{-} \to \mu^{-} \mu^{-} \pi^{+}$ decay}\label{sec:Nmupi}

We start with the calculation of the $B^{-} \to \mu^{-} N$ decay width in \eqref{eq:Btopi_decay}. The Lagrangian terms contributing to the $B^{-} \rightarrow \mu^{-} N$ decay can be explicitly displayed in terms of the massive up and b quark fields in \eqref{eq:mass_basis} as:
\begin{eqnarray}\label{eq:lag_massive}
  \mathcal{L} = \mathcal{L}_{SM} + \frac{1}{\Lambda^2}&&  \Big\{ -\alpha^{(2)}_W \frac{ ~v ~m_W}{\sqrt{2}}\, \overline{\mu} \gamma^{\nu} P_R N  \, W^{-}_{\nu}    
		+  \alpha^{(2)}_{V_0} U_{R}^{12~*} ~D_{R}^{23} \, \overline{u} \gamma^{\nu} P_R b \,  \, \overline{\mu} \gamma_{\nu} P_R N   \nonumber \\
		&& +  \alpha^{(2)}_{S_1} U_{R}^{12~*} ~D_{L}^{23} \, \, \overline{u} P_L b \, \, \overline{\mu} P_R N -  \alpha^{(2)}_{S_2} U_{L}^{12~*} D_{R}^{23} \, \overline{u} P_R b \, \, \overline{\mu} P_R N  \nonumber \\
		&& + \alpha^{(2)}_{S_3}  U_{L}^{12~*} D_{R}^{23} \,\,  \overline{u} P_R N \, \,  \overline{\mu} P_R b \Big\},
\end{eqnarray}
The new quark flavor-mixing matrix element products of $U_{R,L}$ and $D_{R,L}$ can be renamed for simplicity as
\begin{eqnarray}\label{eq:YX_mix_ub} 
Y^{ub}_{RR} \equiv U_{R}^{12~*} ~D_{R}^{23}, \qquad 
Y^{ub}_{RL} \equiv U_{R}^{12~*} ~D_{L}^{23},  \qquad
Y^{ub}_{LR} \equiv U_{L}^{12~*} ~D_{R}^{23}.  
\end{eqnarray}
%
In Appendix \ref{sec:bdec_muN} we show the details of the calculation leading to the decay width $\Gamma^{B\rightarrow \mu N}$. The result is
\begin{eqnarray}\label{eq:BtomuN_width}
\Gamma^{B\rightarrow \mu N}&=&\frac{1}{16\pi m_B}\left(\frac{f_B m_B^2}{2 \Lambda^2} \right)^2 \left\{ |A_V|^2 \left[
(1+B_{\mu}-B_N)(1-B_{\mu}+B_N)-(1-B_{\mu}-B_N)\right] 
\right. 
\nonumber \\
&+& \left.
 |A_S|^2\frac{(1-B_{\mu}-B_N)}{(\sqrt{B_u}+\sqrt{B_b})^2} + (A^*_S A_V+A^*_V A_S) \frac{\sqrt{B_{\mu}}(1-B_{\mu}+B_N)}{(\sqrt{B_u}+\sqrt{B_b})}    \right\}
\nonumber \\
& \times & \sqrt{(1-B_{\mu}+B_N)^2-4 B_N},
\end{eqnarray}
where $B_{\mu}= m_{\mu}^2/m_B^2 , \;\; B_{N}=m_{N}^2/m_B^2,  \;\; B_{u} = m_{u}^2/m_B^2, \;\; B_{b}=m_{b}^2/m_B^2$, and 
\begin{eqnarray}\label{eq:A_couplings}
 A_V &=&  \left( \alpha^{(2)}_{V_0} Y^{ub}_{RR} + \alpha_W^{(2)} V^{ub} \right) \nonumber \\
 A_S &=&  \left( \alpha^{(2)}_{S_1} Y^{ub}_{RL} + (\alpha^{(2)}_{S_2}  +   \frac12 \alpha^{(2)}_{S_3}) Y^{ub}_{LR} \right). 
\end{eqnarray}
The effective couplings in $A_{V,S}$ -as the subscript indicates- correspond to \emph{vectorial} and \emph{scalar} interactions.

Also from the Lagrangian in \eqref{eq:lag_massive}, but changing the $b$ quark fields by $d$ quarks, we find the decay width for the $N\to\pi^+ \mu^-$ process pictured in Fig. \ref{fig:Bmumupi} to be
\begin{eqnarray}\label{eq:Ntopimu_width}
  \Gamma(N\to\pi^+ \mu^-) &=& \frac{1}{16\pi  m_N} \left( \frac{f_{\pi} m_N^2}{2\Lambda^2} \right)^2  \Big\{ |C_V|^2 \left[ (1-P_{\mu} - P_{\pi}) (1 -P_{\mu} +P_{\pi} )-P_{\pi}(1+P_{\mu}-P_{\pi}) \right]   
\nonumber \\   
& - &    (C^*_S C_V + C^*_V C_S) \frac{P_{\pi} \sqrt{P_{\mu}}}{\sqrt{P_{u}} +\sqrt{P_{d}}}(1-P_{\mu}+P_{\pi})
\nonumber \\   
& + &  |C_S|^2 \frac{P_{\pi}^2}{\big(\sqrt{P_u}+\sqrt{P_d}\big)^2} (1+P_{\mu}-P_{\pi})
   \Big\}  \sqrt{ (1-P_{\mu}-P_{\pi} )^2-4P_{\pi}}, 
\end{eqnarray}
where $P_{\mu} = {m_{\mu}^2}/{m_N^2}, \;\; P_{\pi} = {m_{\pi}^2}/{m_N^2}, \;\;  P_u = {m_u^2}/{m_N^2},\;\;   P_d = {m_d^2}/{m_N^2},$ and \footnote{Again, the effective couplings in $C_{V,S}$ correspond to \emph{vectorial} and \emph{scalar} interactions. }
\begin{eqnarray}\label{eq:C_couplings}
 C_V &=& \left( \alpha^{(2)}_{V_0} Y^{ud}_{RR} +\alpha_W^{(2)} V_{ud} \right) \nonumber \\
 C_S &=& \left( \alpha_{S_1}^{(2)} Y^{ud}_{RL} + (\alpha_{S_2}^{(2)}  
       + \frac{1}{2} \alpha^{(2)}_{S_3}) Y^{ud}_{LR} \right).
\end{eqnarray}
The details of the calculation are presented in Appendix \ref{sec:bdec_pion}.

Finally the decay width for the $B^-$ meson $\Gamma(B^-\to \mu^-\mu^- \pi^+)$ is calculated according to \eqref{eq:Btopi_decay} and \eqref{eq:BrNtopi}, allowing us to obtain the effective branching ratio: 
\begin{eqnarray}\label{eq:BrBmumupi_eff}
 Br^{eff}({B^-\to\mu^-\mu^- \pi^+})= \frac{\Gamma(B^-\to \mu^- N)}{\Gamma_B} \frac{\Gamma(N \to \mu^- \pi^+)}{\Gamma_N}. 
\end{eqnarray}
which we compare with the experimental results \cite{Aaij:2014aba}.

\subsection{The $B^- \rightarrow \mu^{-} \nu \gamma $ decay.} \label{sec:bdec_radiative}

The SM radiative leptonic $B$ decays have been extensively studied in the literature  \cite{Beneke:2011nf,Wang:2016qii,DescotesGenon:2002mw,Korchemsky:1999qb,Wang:2018wfj,Beneke:2018wjp}, as they are a means of probing the strong and weak SM interactions in a heavy meson system. The measurement of pure leptonic $B$ decays is very difficult due to helicity suppression and the fact of having only one detected final state particle. On the other hand the radiative modes, with an extra real final photon, can be even larger than the pure leptonic modes as they escape helicity suppression and are also easier to reconstruct. 

 \begin{figure*}[h]
 \centering
  \includegraphics[width=0.8\textwidth]{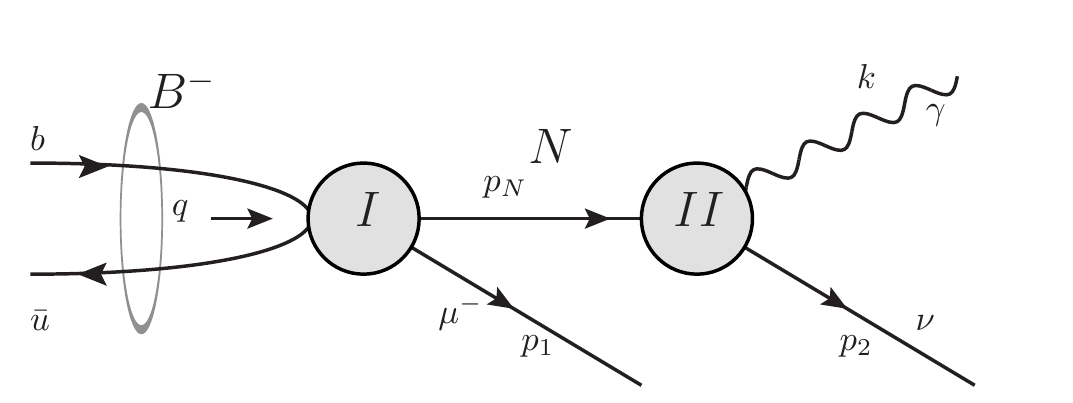}
 \caption{\label{fig:Bmunugamma} Schematic representation for the effective contribution to the decay $B^-\rightarrow  \mu^-\nu \gamma$.}
 \end{figure*}

The Belle collaboration has recently released an analysis of the full Belle experiment dataset \cite{Gelb:2018end} using new theoretical inputs \cite{Beneke:2018wjp} for the QCD calculations and new algorithms prepared for the Belle II experiment. They obtain the experimental bound $\Delta Br^{exp}_{B^- \rightarrow \mu^{-} \nu \gamma}< 3.4 \times 10^{-6}$ for the integrated partial branching ratio of the muon-mode radiative $B$ decay.    

We consider the SM and the effective contribution coming from the $B \rightarrow \mu N$ followed by $N \rightarrow \nu \gamma$ reaction as pictured in Fig. \ref{fig:Bmunugamma}, and use the Belle bound to set limits on the one-loop generated effective couplings involved in this last decay mode, as will be discussed in Sect. \ref{sec:numerical}.

Again the details of the calculations are presented in Appendix \ref{sec:ap_bdec_rad}. There, we obtain the SM value $\Delta Br^{SM} \sim 5 \times 10^{-7} $, which is of the order of the values recently considered in ref. \cite{Zuo:2018sji}. The effective contribution to the ${B^- \rightarrow \mu^{-} \nu \gamma}$ decay is found by integrating the following expression  
\begin{eqnarray}\label{eq:DeltaBr_eff}
\Delta Br^{eff}=\int_{E_{\gamma}^{min}}^{E_{\gamma}^{max}} dE_{\gamma} \frac{d\Gamma^{(B\rightarrow \mu \nu \gamma )}}{dE_{\gamma}}=
\Gamma^{(B\rightarrow \mu N )} Br(N\rightarrow \nu\gamma) \frac{(E_{\gamma}^{max}-E_{\gamma}^{min})}{E_N \beta_N}
\end{eqnarray}
for the allowed range of photon energies, with a minimal infrared cut (to ensure a valid QCD treatment) set to $E_{cut}=1$ GeV, as considered in the Belle experiment.


\section{Results and discussion}\label{sec:results}

\subsection{Numerical treatment}\label{sec:numerical}

The aim of this work is to study the bounds that can be set on the different couplings $\alpha_{\mathcal{J}}$ in the effective dimension 6 Lagrangian \eqref{eq:Lagrangian} involved in $N$ mediated $B$ decays by exploiting the experimental results existing on the  $B^{-} \to \mu^{-} \mu^{-} \pi^{+}$ \cite{Aaij:2014aba} and $B^{-} \to \mu^{-} \nu \gamma$ \cite{Gelb:2018end} processes. 

The numerical value of the couplings $\alpha_{\mathcal{J}}$ can be constrained considering the current experimental bounds on the light-heavy neutrino mixing parameters in low scale minimal seesaw models appearing in the charged interaction in \eqref{eq:mix}. Inspired in this interaction we consider the combination
\begin{equation} \label{eq:u2}
U^2=(\alpha v^2 /(2 \Lambda^2))^2
\end{equation}
which is derived from the contribution of the operator $\mathcal{O}^{(i)}_{LN\phi}$ in \eqref{eq:Ope-SVB} and allows a direct comparison with the mixing angles in the Type I seesaw scenarios \cite{Atre:2009rg}. 

Some of the operators involving the first fermion family (with indices $i=1$) are strongly constrained by the neutrinoless double beta decay bounds, currently obtained by the KamLAND-Zen collaboration \cite{KamLAND-Zen:2016pfg}. Following the treatment already made in \cite{Duarte:2016miz}, the values of the $0\nu \beta \beta$-decay constrained couplings $\alpha^{(1)}_W,\,\alpha^{(1)}_{V_0},\,\alpha^{(1)}_{S_{1,2,3}}$ are taken as equal to the bound $\alpha^b_{0\nu\beta\beta}=3.2\times 10^{-2}\left(\frac{m_N}{100 GeV} \right)^{1/2}$ for $\Lambda=1$ TeV.
These operators appear as contributions to the $\Gamma_N$ width. 

The $B$ to final muon decays studied in this work allow us to set bounds on the couplings involving the second fermion family (generically $\alpha^{(2)}_{\mathcal{J}}$). As we found in Sect. \ref{sec:NmedBdec}, the $B \to \mu^- N$ effective decay depends on the couplings appearing in the vectorial ($A_V$) and scalar ($A_S$) interactions in \eqref{eq:A_couplings}, the $N \to \mu^- \pi^+$ depends on the $C_{V}$ and $C_S$ couplings in \eqref{eq:C_couplings} and the $N \to \nu \gamma$ depends on the one-loop tensorial couplings in \eqref{eq:leff_1loop_L}. 

The new quark flavor-mixing matrices combinations{\footnote{ $Y^{qq'}$ are defined in \eqref{eq:YX_mix_ub} for the $q=u, ~q'=b$ case, and analogously in \eqref{eq:C_couplings} for $q=u, ~q'=d$.}} $Y^{qq'}$ appearing in the first two cases are unknown, and in principle their entries may be found by independent measurements, as is done in the case of the SM $V_{CKM}$ matrix. In this occasion we will make an ansatz and consider that all the $Y^{ub}$ values in \eqref{eq:YX_mix_ub} shall be of the order of the SM $V^{ub}$ value, taking it as a measure of the strength of the coupling between the $u$ and $b$ quarks. Correspondingly, we will consider the $Y^{ud}$ values to be of the order of the SM $V^{ud}$ $CKM$ mixing.

This allows us to consider $A_V$ and $A_S$ in \eqref{eq:BtomuN_width} and  $C_V$ and $C_S$  in \eqref{eq:Ntopimu_width} for the numerical treatment as 
\begin{eqnarray}
 A_V = \left(\alpha^{(2)}_{W} + \alpha^{(2)}_{V_0} \right) V^{ub}   & \qquad&  C_V = \left( \alpha^{(2)}_{W} + \alpha^{(2)}_{V_0} \right) V^{ud} \nonumber \\
 A_S = \left( \alpha^{(2)}_{S_1} + \alpha^{(2)}_{S_2} + \frac12 \alpha^{(2)}_{S_3} \right) V^{ub} & \qquad& C_S = \left( \alpha^{(2)}_{S_1} + \alpha^{(2)}_{S_2} + \frac12 \alpha^{(2)}_{S_3}\right) V^{ud}  
\end{eqnarray}
and set bounds on the possible values of these effective couplings using the $B$-decay data. 

As we would like to disentangle the kind of new physics contributing to the Majorana neutrino interactions, for the numerical analysis we will consider different benchmark scenarios for the effective couplings, where we switch on/off the operators with distinct Dirac-Lorentz structure: vectorial, scalar and the tensorial one-loop generated operators. If we call ($V, S, L$) the factors multiplying the vectorial, scalar and one-loop generated operators respectively, we can define six sets, presented in table \ref{tab:alpha-sets}. 

\begin{table}[t]
 \centering
 \begin{tabular}{|l l c |c c c c c c|}
\firsthline
Operators & Couplings  & $\;$ Type $\;$ & \bf{Set1} &  \bf{Set2}  &  \bf{Set3} & \bf{Set4} &  \bf{Set5} & \bf{ Set6}\\
\hline
$\mathcal{O}_{L N \phi}$, $\mathcal{O}_{duNL}$  & $\alpha^{(2)}_{W}$ $\alpha^{(2)}_{V_0}$  & V & 1 & 1 &0 &1 & 1& 0 \\  
$\mathcal{O}_{QuNL}$, $\mathcal{O}_{LNQd}$, $\mathcal{O}_{QNLd}\qquad$ & $\alpha^{(2)}_{S_1} $ $\alpha^{(2)}_{S_2}$ $\alpha^{(2)}_{S_3}$ & S & 1  & 0 & 1& 1 & 0 & 1   \\
$\mathcal{O}_{NB}$, $\mathcal{O}_{NW}$ & $\alpha^{(2)}_{NB}$ $\alpha^{(2)}_{NW}$  & L & 1 & 1 & 1& 0 & 0  & 0 \\ 
\lasthline
 \end{tabular}
\caption{Effective operators benchmark sets. }\label{tab:alpha-sets}
\end{table}

In order to exploit $B$-decay data to put bounds on the effective couplings in table \ref{tab:alpha-sets}, we will take them as equal to the same value $\alpha$, and use the experimental results constraining the value of the combination $U^2$ defined in \eqref{eq:u2}. We have $\alpha=2\Lambda^2/v^2 \sqrt{U^2}$ for the tree-level generated operators (which are the vectorial and scalar operators), and in the case of one-loop generated operators we have $\alpha= \frac1{16 \pi^2}\frac{2\Lambda^2}{v^2}\sqrt{U^2}$. This allows us to write the numerical results for the total Majorana neutrino decay width $\Gamma_N$, the branching ratio $Br^{eff}({B^-\to\mu^-\mu^- \pi^+})$ in \eqref{eq:BrBmumupi_eff} and the integrated effective branching ratio $\Delta Br^{eff}(B^- \to \mu \nu \gamma)$ in \eqref{eq:DeltaBr_eff} as a function of the Majorana neutrino mass $m_N$ and the $U^2$ combination. 

Sets 1, 2 and 3 in Tab. \ref{tab:alpha-sets} take into account the contributions of the one-loop generated effective couplings in \eqref{eq:leff_1loop_L} to the $N$ decay width. In particular these sets allow for the existence of the $N\to \nu \gamma $ decay channel represented in Fig. \ref{fig:N_decay}. As we found in \cite{Duarte:2016miz}, this channel gives the dominant contribution to the $N$ decay width for the low mass $m_N$ range considered in this work. Sets 4, 5 and 6 discard this contribution. As can be seen in Fig. \ref{fig:ancho}, the total $\Gamma_N$ width is around three orders of magnitude higher in sets 1, 2 and 3 (Fig.\ref{fig:ancho_set123}) than in sets 4, 5 and 6 (Fig.\ref{fig:ancho_set456}). In fact, as the scalar and vectorial couplings contribution to the $N$ decay in this mass range is so poor, the three curves in Fig.\ref{fig:ancho_set123} cannot be distinguished in the plot scale. This effect in the $\Gamma_N$ value will explain many of the differences in the bounds we obtain for the $U^2$ combination when we consider one group of sets or the other, as will be discussed below.

\subsection{Obtained bounds}\label{sec:obt_bounds} 

We start by discussing the bounds obtained from the LHCb results on the $B^-\rightarrow  \mu^-\mu^- \pi^+$ decays. The LHCb collaboration has presented a search for Majorana-mediated $B^-\rightarrow  \mu^-\mu^- \pi^+$ decays \cite{Aaij:2014aba}, where they obtain model independent limits on the branching ratio $Br(B^-\to \mu^- \mu^- \pi^{+})= Br(B^-\to \mu^- N) . Br(N \to \mu^- \pi^{+})$ as a function of the Majorana mass $m_N$ and lifetime $\tau_N$, ranging from $1$ to $1000$ ps. The results are presented in their figure 5, where they show the upper limits obtained for the above product, at $95\%$C.L.

 \begin{figure*}[h]
 \centering
  \includegraphics[width=0.7\textwidth]{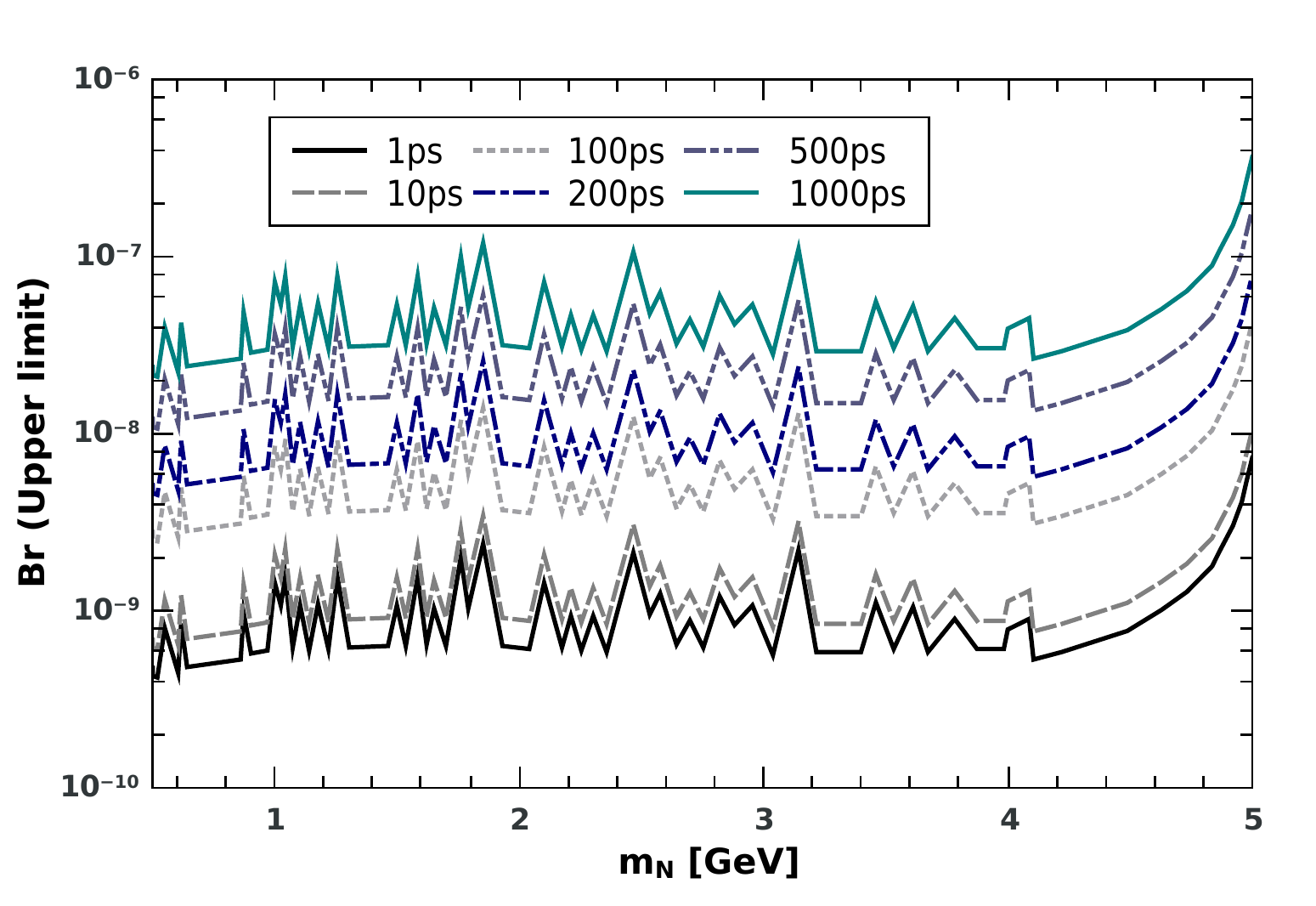}
 \caption{\label{fig:datos} Data on $Br(B^-\to \mu^- \mu^- \pi^{+})$ reproduced from Fig.5 in \cite{Aaij:2014aba}.}
 \end{figure*}

\begin{figure*}
\centering
\subfloat[Sets 1, 2, 3]{\label{fig:set123}\includegraphics[totalheight=6.5cm]{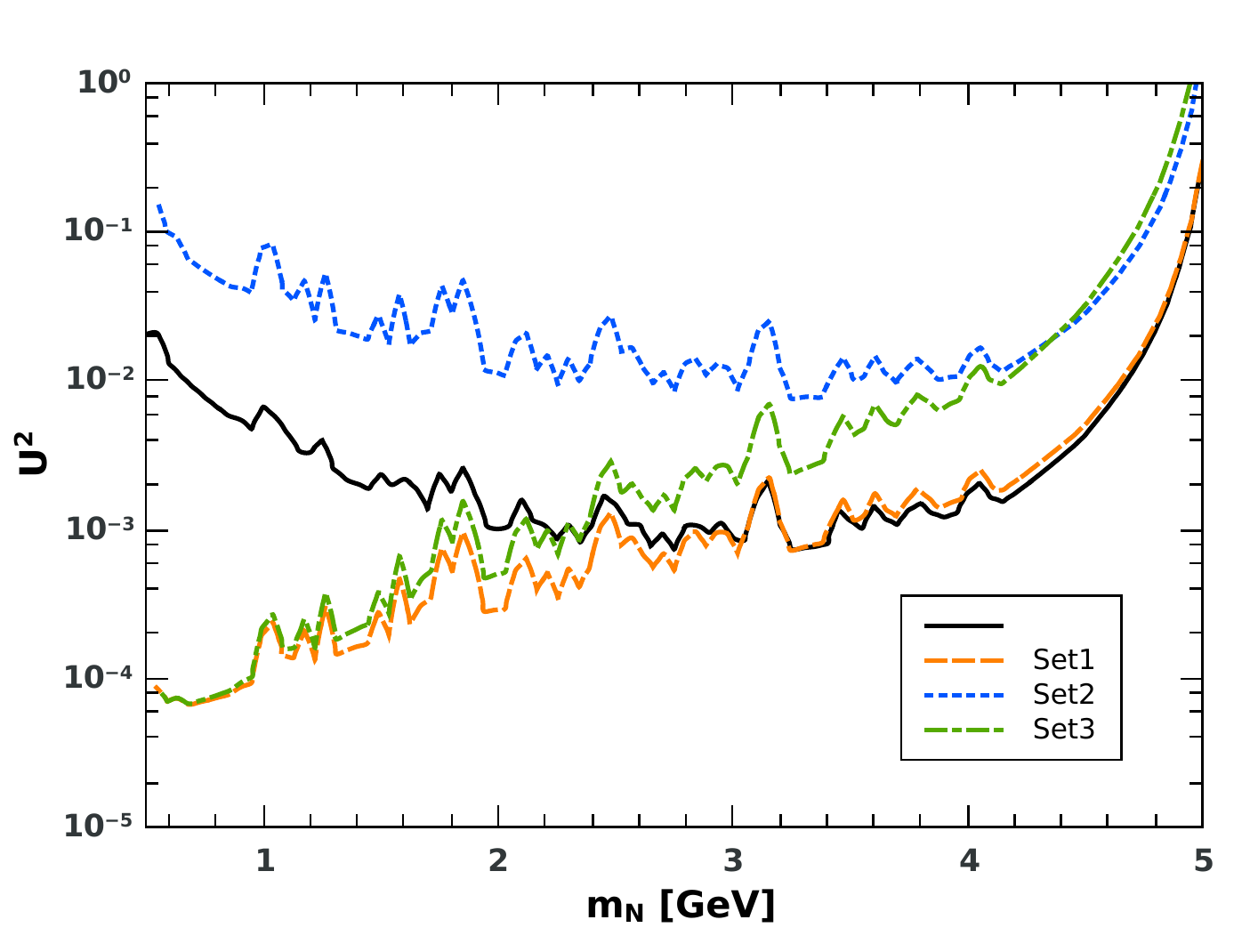}}~
\subfloat[Sets 4, 5, 6]{\label{fig:set456}\includegraphics[totalheight=6.5cm]{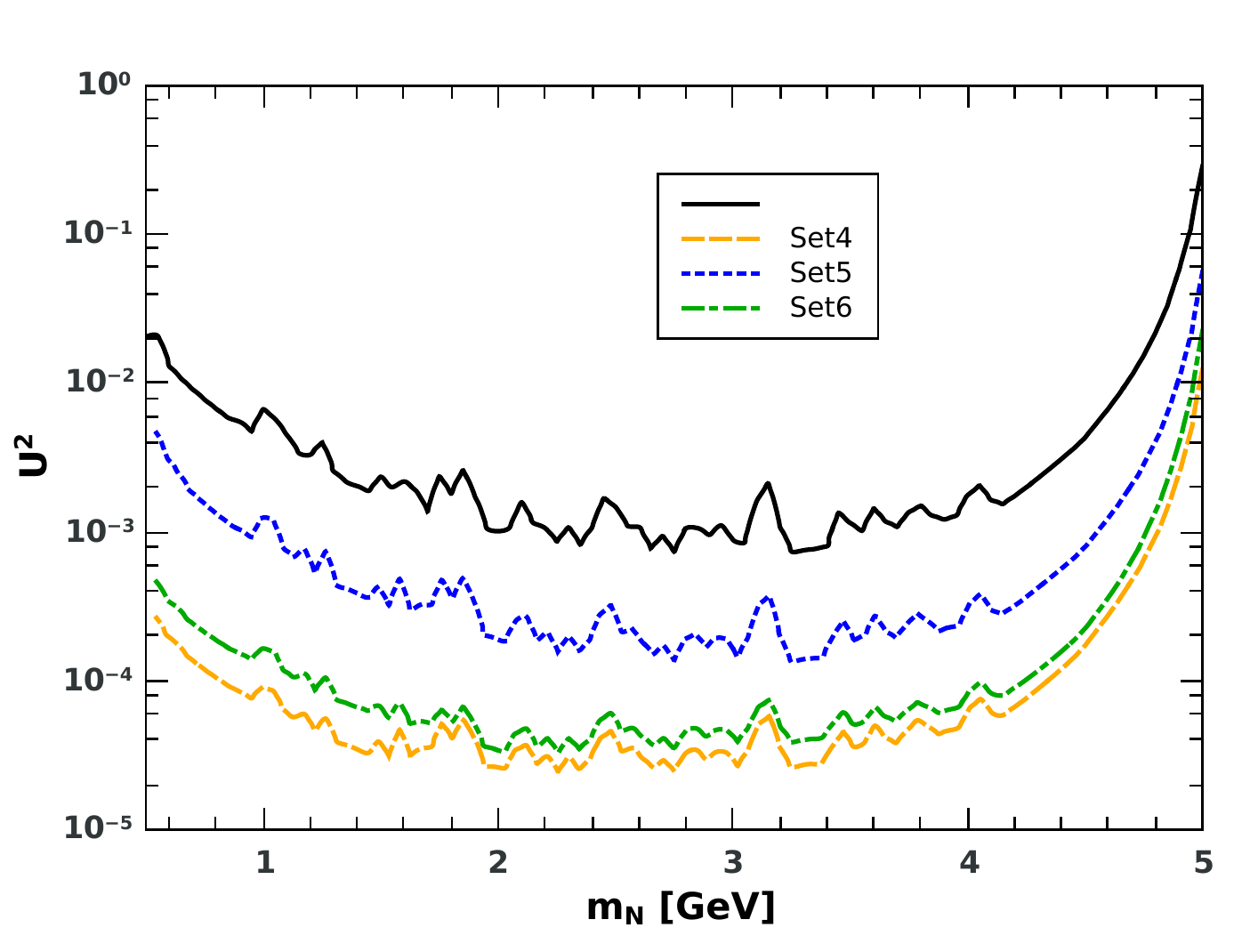}}
\caption{Upper bounds obtained for $U^2$ from $B^-\to \mu^- \mu^- \pi^{+}$ decays\cite{Aaij:2014aba}, considering the effective coupling sets defined in Table \ref{tab:alpha-sets}. The black full line curve represents the revised bound presented in \cite{Shuve:2016muy} for $|U_{\mu N}|^2$.}\label{fig:Bmumupi_bounds}
\end{figure*}

Following the procedure in ref. \cite{Shuve:2016muy}, we convert the model independent LHCb upper limits on the branching ratio $Br(B^-\to \mu^- \mu^- \pi^{+})$ into limits on the combination $U^{2}$ defined in \eqref{eq:u2}. For each value of $m_N$ (which fixes the value of $\tau_N$ for a given $U^2$ value in the effective model) we scan through the values of $U^{2}$ for which our computed branching fraction \eqref{eq:BrBmumupi_eff} equals the upper bound in ref. \cite{Aaij:2014aba}. For the experimental values, we consider the data in Fig. \ref{fig:datos}, which reproduces the values presented in Figure 5 of ref. \cite{Aaij:2014aba}. The obtained constraints on the $U^{2}$ values are presented in Fig. \ref{fig:set123} for the coupling sets 1, 2 and 3 in Tab. \ref{tab:alpha-sets}, and  Fig. \ref{fig:set456} for sets 4, 5 and 6. For comparison, in both figures we add the revised bounds obtained by the authors in ref. \cite{Shuve:2016muy}, where they derive upper bounds on the Type-I seesaw mixing angle $|U_{\mu N}|^2$ in \eqref{eq:mix} from the LHCb results. 

As can be seen in the plots, the bounds we obtain for $U^2$ in sets 1, 2 and 3 (Fig. \ref{fig:set123}) are weaker than those we get in sets 4, 5 and 6 (Fig. \ref{fig:set456}). This is explained by the different $\Gamma_N$ values in the two groups of sets discussed above: in the case of sets 4, 5 and 6, the value of the branching ratio $Br^{eff}({B^-\to\mu^-\mu^- \pi^+})$  in \eqref{eq:BrBmumupi_eff} is around $10^3$ times higher, because the $\Gamma_N$ factor in the denominator is lower than in sets 1, 2 and 3, and thus we obtain more restrictive bounds for the $U^2$ combination when we do not take into account the one-loop generated effective couplings contribution.

On the other hand, among the sets in each figure, we find that we place stronger bounds on the scalar couplings (considering their sole effect in set 6, and with one-loop couplings in set 3). This is due to the presence of the light quark masses in the denominators of the scalar terms \eqref{eq:pseudoescalar_pi}, enhancing these contributions to the $Br(B^- \to \mu^- N)$ in \eqref{eq:BtomuN_width}  and the $Br(N \to \mu^- \pi^-)$ in \eqref{eq:Ntopimu_width}. In sets 1 and 4 the contributions of the scalar and vectorial couplings are considered together. 

Regarding the behavior of the curves for Majorana masses $m_N$ near the $B$ mass, in this limit the decay $B \to \mu N$ is kinematically suppressed. This of course loosens the constraint on $U^2$, as can be seen in both Figs. \ref{fig:set123} and \ref{fig:set456}. A more detailed analysis of the decay width $\Gamma(B \to \mu N)$ in \eqref{eq:BtomuN_width} shows that the third term in the right hand side, involving the product of vectorial and scalar operators, is the one going to zero more slowly in this limit: while the sets 2 and 3 both give a null contribution from this term, as they put scalar or vectorial operators equal to zero, set 1 keeps this contribution. This is why the curves for sets 2 and 3 in Fig. \ref{fig:set123} show similar behaviors, while the curve for set 1 still presents a stronger constraint. The effect is not as big in the curves of Fig. \ref{fig:set456}, given that in these sets the different value of $\Gamma_N$, as shown in Fig. \ref{fig:ancho_set456} separates the curves for sets 5 and 6. We have checked that the apparent close matching of the curves in Fig. \ref{fig:set123} for set 1 and the constraint found by the authors in ref. \cite{Shuve:2016muy} does not originate with any particular physics effect, but is a numerical coincidence caused by the limited data number and the finite precision in the comparison with the theoretical prediction.\\

Now we present the bounds imposed on $U^2$ by the Belle result on the radiative $B\rightarrow \mu \nu \gamma$ decay. 
We compare the results obtained in Sect. \ref{sec:bdec_radiative} for the expressions of the integrated branching fractions for the $B\rightarrow \mu \nu \gamma$ decay: the SM $\Delta Br^{SM} \sim 5 \times 10^{-7} $ in \eqref{eq:Delta_Br_SM} and the effective $\Delta Br^{eff}$ in \eqref{eq:DeltaBr_eff}, with the Belle result \cite{Gelb:2018end}.  

\begin{figure*}
 \centering
  \includegraphics[width=0.7\textwidth]{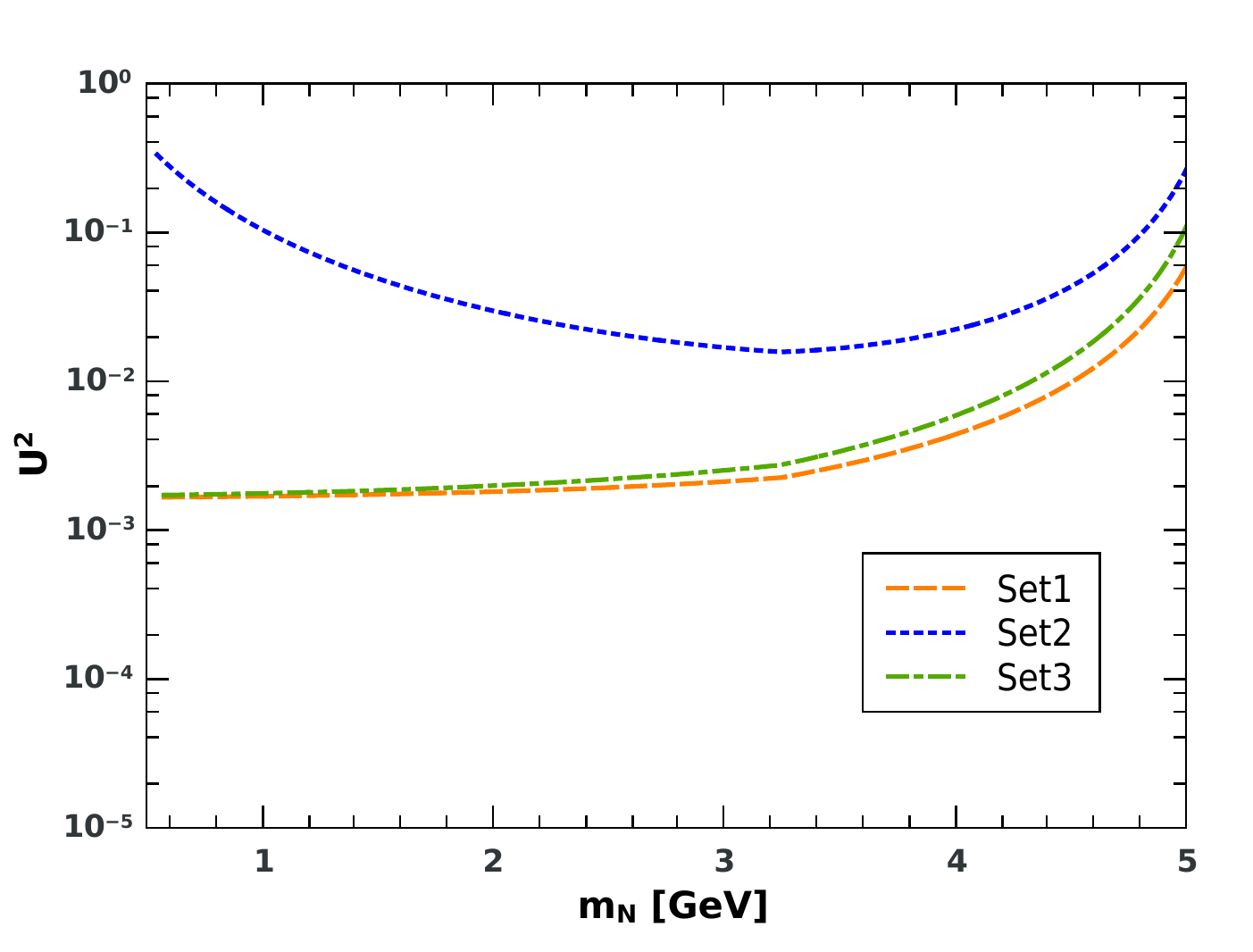}
 \caption{\label{fig:bounds_Brad} Upper bounds obtained for $U^2$ from $B\rightarrow \mu \nu \gamma$ decays \cite{Gelb:2018end}.}
 \end{figure*}

We now scan for the values of $U^2$ for which the complete theoretical value $\Delta Br = \Delta Br^{SM} + \Delta Br^{eff}$ equals the upper bound $\Delta Br^{exp}_{B^-\to \mu^{-} \nu \gamma}< 3.4 \times 10^{-6}$, for each mass $m_N$. The bounds we obtain for $U^2$ from this procedure are presented in Fig. \ref{fig:bounds_Brad}. 
As the one-loop generated operators need to be non-zero to allow the $B^- \rightarrow \mu^- \nu \gamma$ decay, bounds are established just for the sets 1, 2 and 3. Again, we obtain stronger bounds on the scalar operators, due to their contribution to the $N$ production in $B$ decay \eqref{eq:BtomuN_width}. These bounds are compatible with and less restrictive than the ones obtained from the LNV process $B^-\to \mu^- \mu^- \pi^+$ in Fig. \ref{fig:set123}. \\

While the sensibility to effective interactions looks hard to improve for the case of the $B\to \mu \mu \pi$ decay measurements, we find there is room for new analyses concerning the $B \to \mu \nu \gamma$ decay, specially in $B$ factories. One can think of resonance searches using the $B$ mass and the $e^+ e^-$ beam energy to constrain the missing momentum and infer the $N$ mass. Also, in this low $m_N$ range, as we found in \cite{Duarte:2016miz, Duarte:2015iba}, the Majorana neutrino is long-lived, and it would be possible to search for displaced photons together with a prompt lepton and missing $E_T$ in the final state, a possibility we already explored for the LHC $pp\to \mu \nu \gamma$ reaction in \cite{Duarte:2016caz}. Other observables involving the final photon or charged lepton polarizations could improve also the sensitivity to the different contributions from vectorial and scalar operators. As we recently explored in \cite{Duarte:2018kiv}, the study of final tau polarizations for the $B \to \tau \nu \gamma$ decay could help disentangle vectorial and scalar contributions in the intermediate production process $B \to N \tau$.

For comparison with earlier work, we take the minimum values for $U^2$ from Fig. \ref{fig:Bmumupi_bounds} in each set, and calculate the maximum allowed $\alpha$ value, considering the new physics scale to be $\Lambda=1$ TeV. The bounds obtained in this way are shown in Table \ref{tab:alpha-bounds}. 
The lower $\alpha \leq 0.16$ value is found (for $m_N=2.26$ GeV) in the coupling set 4, which considers vectorial and scalar interactions. It grows to $\alpha \leq 0.19$ when only scalar interactions are included (set 6). When one-loop generated interactions are also taken into account, the bound is relaxed to $\alpha \leq 0.26$ in sets 1 and 3.    

The $\alpha \leq 0.16$ bound should be compared for instance with the upper bound our group considered for the calculation of the contribution of scalar and vectorial effective Majorana interactions to the LNV same-sign dilepton signal $pp \to \mu^{+} \mu^{+} jj$ in the LHC \cite{Duarte:2016caz}. In that early work we estimated an upper bound $\alpha \leq 0.3$ coming from the heavy neutrinos search results at Belle \cite{Liventsev:2013zz}. Other works also took into account the same bound for the calculation of prospects for the observation of $e^{+}e^{-} \to  \nu N \to \nu \gamma$ at Belle-II and the ILC \cite{Yue:2017mmi}. The revision of these results is left for future work. 

As we mentioned in the introduction, the effective dimension 6 operators parameterize a wide variety of UV-complete models which introduce new degrees of freedom, as the Left-Right symmetric model (LRSM). In ref. \cite{Ruiz:2017nip} bounds are obtained on the couplings of the effective vector  $\mathcal{O}_{duNl}=\mathcal{O}_{V_0}$  and scalar $\mathcal{O}_{QNLd}=\mathcal{O}_{S_3}$  four-fermion contact operators in \eqref{eq:Ope-4-f}. The result is obtained performing a reinterpretation in terms of the LRSM model of the LHC limits from heavy Majorana neutrino direct production at $\sqrt{s}=8$ TeV in the dilepton channel $pp \to W^{*}_{R} \to N \mu^{\pm} \to \mu^{\pm} \mu^{\pm} +nj$ \cite{Khachatryan:2015gha}.   
The most stringent bounds on $\alpha_{V_{0}, S_{3}}$ are obtained considering $Br(N\to \mu X) \sim 1$, so that the $N$ decays preferably to muons. These are (taking $\Lambda=1$ TeV for comparison) $\alpha_{V_{0}} \lesssim 0.23$  and $\alpha_{S_{3}} \lesssim 0.45$ for $m_N=100$ GeV. 

The comparison suggests that the direct calculation of the effective $N$ interactions contribution to different processes can help to put more stringent bounds to different UV-complete models parameterized by the effective Lagrangian in \eqref{eq:Lagrangian}.

\begin{table}[t]
 \centering
 \begin{tabular}{| c |c | c | c | c | c | c|}
\firsthline
 & \bf{Set1} &  \bf{Set2}  &  \bf{Set3} & \bf{Set4} &  \bf{Set5} & \bf{ Set6}\\
\hline
$U^2$  & $6.7\times 10^{-5}$  & $7.7\times 10^{-3}$  & $6.7\times 10^{-5}$ & $2.5\times 10^{-5}$  & $1.3 \times 10^{-4}$  & $3.4 \times 10^{-5}$  \\  
 $\alpha$ & $0.26$ &  $2.8$ & $0.26$  & $0.16$ & $0.36$ & $0.19$   \\
\lasthline
 \end{tabular}
\caption{The most stringent bounds on the tree-level effective couplings $\alpha$ from Fig. \ref{fig:Bmumupi_bounds}, for $\Lambda=1$ TeV. }\label{tab:alpha-bounds}
\end{table}

\section{Final remarks} \label{sec:final}

We have considered heavy Majorana neutrinos coupled to the ordinary matter in a general way by dimension 6 effective operators satisfying the SM electroweak symmetry. According to these interactions these neutrinos would be produced in the decay of $B$ mesons, and subsequently decay to standard particles. In particular, we exploited the non-observation of the $B^- \rightarrow \mu^- \mu^- \pi^+ $ decay in the LHCb \cite{Aaij:2014aba} and put limits to the couplings of the different effective operators contributing to this decay in the Majorana mass range $0.5$ GeV $<m_N<5$ GeV. These upper bounds are presented in Fig. \ref{fig:Bmumupi_bounds}.  

Also for this $m_N$ range, we have considered the bounds coming from the radiative decay $B^- \rightarrow \mu^- \nu \gamma $ by the Belle experiment \cite{Gelb:2018end}. This allows us to set bounds directly on the one-loop generated operators. These bounds are compatible with and weaker than the ones we derive form the LNV process $B^-\rightarrow \mu^-\mu^-\pi^+$ and are shown in Fig. \ref{fig:bounds_Brad}. 

The obtained bounds (for $0.5 \lesssim m_N \lesssim 5 $ GeV) are more restrictive than previous values obtained for dimension 6 four-fermion contact vectorial and scalar Majorana neutrino interactions in the context of the Left-Right symmetric model for higher Majorana masses \cite{Ruiz:2017nip}. The comparison suggests that the direct calculation of the effective $N$ interactions contribution to different processes can help to put more stringent bounds to different UV-complete models parameterized by the effective interaction formalism. The obtained upper bounds  also constrain the perspective of discovery of Majorana neutrinos with GeV-scale masses by direct production in colliders and meson decays \cite{Duarte:2016caz, Yue:2017mmi, Yue:2018hci}.

{\bf{Note added:}} While this manuscript was under revision, two works appeared concerning the study of effective interactions involving right handed neutrinos, \cite{Bischer:2019ttk} and \cite{Alcaide:2019pnf}, where bounds are obtained for some effective operators.

\appendix

\section{The $B^- \rightarrow \mu^{-} N $ decay.} \label{sec:bdec_muN}

From the Lagrangian in \eqref{eq:lag_massive},  we find the amplitude for the process $B^{-}\rightarrow \mu^- N$ is 
\begin{eqnarray}\label{eq:M_BmuN}
 \mathcal{M}_{(B^{-}\rightarrow \mu^- N)}=\langle N\mu^-| \mathcal{L} |B\rangle &=& \frac{1}{\Lambda^2}  
 \left\{ -\alpha_W^{(2)} V^{ub} \left<0\vert \bar u \gamma^{\nu} P_L b \vert B \right>  \left<N \mu \vert \bar \mu \gamma_{\nu} P_R N \vert 0\right>  \right.
\nonumber \\ 
 &&     \left. + \alpha^{(2)}_{V_0} Y^{ub}_{RR}  \left<0\vert \bar u \gamma^{\nu} P_R b \vert B \right>  \left<N \mu \vert \bar \mu \gamma_{\nu} P_R N \vert 0\right> \right.
\nonumber \\
&&\left. 
+\alpha^{(2)}_{S_1} Y^{ub}_{RL} \left<0\vert \bar u P_L b \vert B \right>  \left<N \mu \vert \bar \mu P_R N \vert 0\right> \right.
\nonumber \\
&&\left. 
-\alpha^{(2)}_{S_2} Y^{ub}_{LR} \left<0\vert \bar u P_R b \vert B \right>  \left<N \mu \vert \bar \mu P_R N \vert 0\right>
\right.
\nonumber \\
&&\left. 
+ \alpha^{(2)}_{S_3} Y^{ub}_{LR} \left<N \mu \vert \bar u P_R N \, \bar \mu P_R b \vert B\right> \right\},
\end{eqnarray}
The first term in the amplitude corresponds to the $W$-mediated diagram which includes a SM vertex, giving the $CKM$ $V^{ub}$ contribution.  
In the last term, we need to rearrange the field operators in order to put together the quark fields in a sandwich and the lepton fields in another. So we make a Fierz transformation taking into account a minus sign from the permutation of fermions, and then we replace it by 
\begin{eqnarray*}
- \frac12 \alpha^{(2)}_{S_3} Y^{ub}_{LR}  \left[ \left<0\vert \bar u P_R b \vert B \right>  \left<N \mu \vert \bar \mu P_R N \vert 0\right> +  \frac12 \left<0\vert \bar u \sigma^{\mu\nu}P_R b \vert B \right>  \left<N \mu \vert \bar \mu \sigma_{\mu\nu} P_R N \vert 0\right>\right].
\end{eqnarray*}
The calculation of the leptonic matrix elements is straightforward, 
\begin{eqnarray}
\left<N \mu \vert \bar \mu \gamma_{\nu} P_R N \vert \; 0 \; \right> &=& \bar u_{\mu}(p_1) \gamma_{\nu} P_R v_N(p_N)
\nonumber \\
\left<N \mu \vert \bar \mu P_R N \vert \; 0 \; \right> &=& \bar u_{\mu}(p_1) P_R v_N(p_N)
\end{eqnarray}
In order to calculate the hadronic matrix elements, we have to rely on the symmetries \cite{Campbell:2008um, Shanker:1982nd}. The matrix element 
$\left<\;0\vert \bar u \gamma^{\nu} \gamma_5 b \vert \; B \; \right>$ is a Lorentz 4-vector because the $B$ meson is a pseudoscalar and 
$\bar u \gamma^{\nu} \gamma_5 b$ is a pseudo 4-vector. 
The meson state is described solely by its four momentum $q^{\mu}$, since it has zero spin. Therefore, $q^{\mu}$ is the only 
4-moment on which the matrix element depends and it must be proportional to $q^{\mu}$. Thus, we can write
\begin{eqnarray}
\left<\;0\vert \bar u \gamma^{\nu} \gamma_5 b \vert \; B \; \right>  \;\; = \;\; i f_B q^{\nu}.
\end{eqnarray} 
On the other hand, for the same reason, the matrix elements of the 4-vector, the tensor and pseudo-tensor are zero 
\begin{eqnarray}
\left<\;0\vert \bar u \gamma^{\nu} b \vert \; B \; \right>  \;\; &=& \;\; 0, \nonumber \\
\left<\;0\vert \bar u \sigma_{\mu\nu} b \vert \; B \; \right>  \;\; &=& \;\; 0,
\nonumber \\
\left<\;0\vert \bar u \sigma_{\mu\nu} \gamma_5 b \vert \; B \; \right>  \;\; &=& \;\; 0. 
\end{eqnarray} 
In the case of the matrix element of the scalar or pseudo-scalar interactions, we have to use the Dirac equations of motion, and we obtain the relations for the current matrix elements
\begin{eqnarray}
\left<\;0\vert \bar u \gamma_5 b \vert \; B \; \right>  \;\; &=& \;\; -i \frac{m_B^2 f_B}{m_b+m_u}
\label{eq:pseudoescalar_B} \\
\left<\;0\vert \bar u  b \vert \; B \; \right>  \;\; &=& \;\; 0. 
\end{eqnarray}

Putting it all together and integrating over the 2-body phase space, we obtain 
\begin{eqnarray}
\Gamma^{B\rightarrow \mu N}=\frac{\vert \mathcal{M} \vert^2}{16\pi m_B^3}\sqrt{(m_B^2+m_N^2-m_{\mu}^2)^2-4m_B^2m_N^2},
\end{eqnarray}
with $\mathcal{M}$ in \eqref{eq:M_BmuN} giving 
\begin{eqnarray}\label{eq:M_BmuN2}
 \vert \mathcal{M}_{(B^{-}\rightarrow \mu^- N)} \vert^2 &=& \left(\frac{f_B m_B^2}{2 \Lambda^2} \right)^2 \left\{ \vert A_V \vert^2 \left[ (1+B_{\mu}-B_N) (1-B_{\mu}+B_N)- (1-B_{\mu}-B_N) \right]  \right. \nonumber \\
& &  \left. + \frac{m_{\mu}}{(m_b+m_u)}  (A^{*}_S A_V + A^{*}_V A_S) (1-B_{\mu}+B_N)  \right. \nonumber \\
& & \left. + \frac{m^2_{B}}{(m_b+m_u)^2} \vert A_S \vert^2 (1-B_{\mu}-B_N) \right\}
\end{eqnarray}
where
\begin{eqnarray}
 A_V &=&  \left( \alpha^{(2)}_{V_0} Y^{ub}_{RR} + \alpha_W^{(2)} V^{ub} \right) \nonumber \\
 A_S &=&  \left( \alpha^{(2)}_{S_1} Y^{ub}_{RL} + (\alpha^{(2)}_{S_2}  +   \frac12 \alpha^{(2)}_{S_3}) Y^{ub}_{LR} \right). 
\end{eqnarray}
The result is
\begin{eqnarray}
\Gamma^{B\rightarrow \mu N}&=&\frac{1}{16\pi m_B}\left(\frac{f_B m_B^2}{2 \Lambda^2} \right)^2 \left\{ |A_V|^2 \left[
(1+B_{\mu}-B_N)(1-B_{\mu}+B_N)-(1-B_{\mu}-B_N)\right] 
\right. 
\nonumber \\
&+& \left.
 |A_S|^2\frac{(1-B_{\mu}-B_N)}{(\sqrt{B_u}+\sqrt{B_b})^2} + (A^*_S A_V+A^*_V A_S) \frac{\sqrt{B_{\mu}}(1-B_{\mu}+B_N)}{(\sqrt{B_u}+\sqrt{B_b})}    \right\}
\nonumber \\
& \times & \sqrt{(1-B_{\mu}+B_N)^2-4 B_N},
\end{eqnarray}
where $B_{\mu}= m_{\mu}^2/m_B^2 , \;\; B_{N}=m_{N}^2/m_B^2,  \;\; B_{u} = m_{u}^2/m_B^2, \;\; B_{b}=m_{b}^2/m_B^2$.

\section{The $B^- \rightarrow \mu^{-} \mu^{-} \pi^{+} $ decay.} \label{sec:bdec_pion}

Let us now calculate the decay $N\to\pi^+\mu^-$. According to the Lagrangian \eqref{eq:efflag} written in terms of the massive $u$ and $d$ quarks as we did in \eqref{eq:lag_massive}, the amplitude for this process can be written as
\begin{eqnarray}
  \mathcal{M}_{(N\to\pi^+ \mu^-)} &=& \langle\pi^+\mu^-| \mathcal{L} |N\rangle  \nonumber\\
	    &=& \frac{1}{\Lambda^2}  \Big\{   - \alpha_W^{(2)} V_{ud}\langle \pi^+| \overline{u}\gamma^{\nu}P_L d |0\rangle \,\langle\mu^-| \overline{\mu}\gamma_{\nu}P_R N |N\rangle   
\nonumber\\
&&	\qquad\quad +\alpha^{(2)}_{V_0}  Y^{ud}_{RR} \langle\pi^+| \overline{u}\gamma^{\nu}P_R d |0\rangle \, 
\langle\mu^-| \overline{\mu}\gamma_{\nu}P_R N |N\rangle   
\nonumber\\ 
&& \qquad\quad +\alpha_{S_1}^{(2)} Y^{ud}_{RL} \langle\pi^+| \overline{u}P_L d |0\rangle \, \langle\mu^-| \overline{\mu}P_R N |N\rangle   
\nonumber\\ 
&& \qquad\quad -\alpha_{S_2}^{(2)} Y^{ud}_{LR} \langle\pi^+| \overline{u}P_R d |0\rangle \, \langle\mu^-| \overline{\mu}P_R N |N\rangle
\nonumber\\
&&	\qquad\quad +\alpha_{S_3}^{(2)} Y^{ud}_{LR} \langle\pi^+\mu^-| \overline{u}P_R N\mu^-P_R d |N\rangle	\quad\Big\}.	
							\label{eq:M_Ntopimu}							
\end{eqnarray}
where we have defined the flavor-mixing matrix products $Y^{ud}$ in analogy with \eqref{eq:YX_mix_ub}.

The last term in \eqref{eq:M_Ntopimu} also needs to be  modified by means of a Fierz transformation. After some algebra, it is written as
\begin{equation*}
  \langle\pi^+\mu^-| \overline{u}P_R N\mu^-P_R d |N\rangle  =  
	     -\frac{1}{2} \langle\pi^+| \overline{u}P_R d |0\rangle \,  \langle\mu^-| \overline{\mu}P_R N |N\rangle .
\end{equation*}
In order to calculate the various factors in \eqref{eq:M_Ntopimu}, we make use of the definition for the pion form factor
\begin{equation}
  \langle\pi^+| \overline{u}\gamma^{\mu}\gamma_5 d |0\rangle = ik^{\mu}f_{\pi},
\end{equation}
and from this equation we obtain the following expressions
\begin{equation}
  \langle\pi^+| \overline{u}\gamma^{\nu}P_R d |0\rangle = \frac{i}{2}k^{\nu}f_{\pi},  \qquad
	\langle\pi^+| \overline{u}\gamma^{\nu}P_L d |0\rangle = -\frac{i}{2}k^{\nu}f_{\pi}  \nonumber
\end{equation}
\begin{equation}
  \qquad\mbox{and}\qquad
	\langle\pi^+| \overline{u}P_{R, L} d |0\rangle = \frac{ \pm i}{2}\frac{m^2_{\pi}}{m_u+m_d}f_{\pi}. \label{eq:pseudoescalar_pi}
\end{equation}
The contribution of the pseudo-scalar quark current to the matrix element of the ordinary pion decay \eqref{eq:pseudoescalar_pi} is enhanced in comparison with the standard chirality suppressed $V-A$ contribution and it is expected to be severely constrained by the experimental data.   
We finally have for the squared amplitude
\begin{eqnarray}
  |\mathcal{M}_{(N\to\pi^+ \mu^-)}|^2 &=& \left( \frac{f_{\pi} m_N^2}{2\Lambda^2} \right)^2  \Big\{ |C_V|^2 \left[ (1-P_{\mu} - P_{\pi}) (1 -P_{\mu} +P_{\pi} )-P_{\pi}(1+P_{\mu}-P_{\pi}) \right]   
\nonumber \\   
& - &    (C^*_S C_V + C^*_V C_S) \frac{P_{\pi} \sqrt{P_{\mu}}}{\sqrt{P_{u}} +\sqrt{P_{d}}}(1-P_{\mu}+P_{\pi})
\nonumber \\   
& + &  |C_S|^2 \frac{P_{\pi}^2}{\big(\sqrt{P_u}+\sqrt{P_d}\big)^2} (1-P_{\mu}-P_{\pi})
    \Big\},
\end{eqnarray}
where 
\begin{eqnarray}
 C_V &=& \left( \alpha^{(2)}_{V_0} Y^{ud}_{RR} +\alpha_W^{(2)} V_{ud} \right) \nonumber \\
 C_S &=& \left( \alpha_{S_1}^{(2)} Y^{ud}_{RL} + (\alpha_{S_2}^{(2)}  
       + \frac{1}{2} \alpha^{(2)}_{S_3}) Y^{ud}_{LR} \right).
\end{eqnarray}
from which we obtain the decay width for $N\to\pi^+ \mu^-$
\begin{eqnarray}
  \Gamma(N\to\pi^+ \mu^-) &=& \frac{1}{16\pi  m_N} \left( \frac{f_{\pi} m_N^2}{2\Lambda^2} \right)^2  \Big\{ |C_V|^2 \left[ (1-P_{\mu} - P_{\pi}) (1 -P_{\mu} +P_{\pi} )-P_{\pi}(1+P_{\mu}-P_{\pi}) \right]   
\nonumber \\   
& - &    (C^*_S C_V + C^*_V C_S) \frac{P_{\pi} \sqrt{P_{\mu}}}{\sqrt{P_{u}} +\sqrt{P_{d}}}(1-P_{\mu}+P_{\pi})
\nonumber \\   
& + &  |C_S|^2 \frac{P_{\pi}^2}{\big(\sqrt{P_u}+\sqrt{P_d}\big)^2} (1+P_{\mu}-P_{\pi})
   \Big\}  \sqrt{ (1-P_{\mu}-P_{\pi} )^2-4P_{\pi}} 
\end{eqnarray}
where
\begin{equation*}
   P_{\mu} = \frac{m_{\mu}^2}{m_N^2}, \qquad  P_{\pi} = \frac{m_{\pi}^2}{m_N^2}, \qquad  P_u = \frac{m_u^2}{m_N^2} \qquad   \mbox{and} \qquad   P_d = \frac{m_d^2}{m_N^2}.
\end{equation*}

\section{The $B^- \rightarrow \mu^{-} \nu \gamma $ decay.} \label{sec:ap_bdec_rad}

The SM $B \to \mu \nu \gamma$ differential decay width in the $B$ meson rest frame can be parameterized as \cite{Wang:2018wfj}
\begin{eqnarray}
\frac{d\Gamma^{(B \rightarrow \mu \nu \gamma)}}{dE_{\gamma}}=\frac{\alpha_{emg}G_F^2 \vert V^{ub} \vert^2}{6\pi^2} m_B
E_{\gamma}^3 \left(1- \frac{2 E_{\gamma}}{m_B} \right)\left(F_V(E_{\gamma})^2+F_A(E_{\gamma})^2 \right)
\end{eqnarray}
with form factors $F_{V,A}$ depending on the final photon energy $E_{\gamma}$. Here $\alpha_{emg}$ and $G_F$ are the fine structure and Fermi couplings. In order to perform the energy integration, we estimate the values for the $F_{V,A}$ form factors taking the central values presented in figure (8) of reference \cite{Wang:2018wfj} \footnote{These values are also consistent with the central values given in figures (7) and (8) of reference \cite{Beneke:2018wjp}, for the inverse moment of the leading twist light cone distribution amplitude $\lambda_B$ value given by Belle \cite{Gelb:2018end}.}.     

We call $\Delta Br^{SM}$ to the integrated partial branching ratio in the energy range  $E_{cut}<E_{\gamma}<E_{\gamma}^{max}$,
\begin{eqnarray}\label{eq:Delta_Br_SM}
\Delta Br^{SM}=\frac1{\Gamma_B} \int_{E_{cut}}^{E_{\gamma}^{max}} dE_{\gamma} \frac{d\Gamma^{(B \rightarrow \mu \nu \gamma)}}{dE_{\gamma}}.
\end{eqnarray}
Here for kinematic reasons $E_{\gamma}^{max}=m_B/2$ and the minimal photon energy infrared cutoff $E_{cut}$ is such that the theoretical QCD treatment remains valid. As we will use the latest Belle results for the experimental limit $\Delta Br^{exp}$, we take $E_{cut}=1$ GeV, as in ref. \cite{Gelb:2018end}.
The value we obtain for our estimation of the partial branching ratio in the SM is $\Delta Br^{SM} \sim 5 \times 10^{-7} $, which is of the order of the values recently considered in ref. \cite{Zuo:2018sji}.\\

\begin{figure*}
 \centering
  \includegraphics[width=0.7\textwidth]{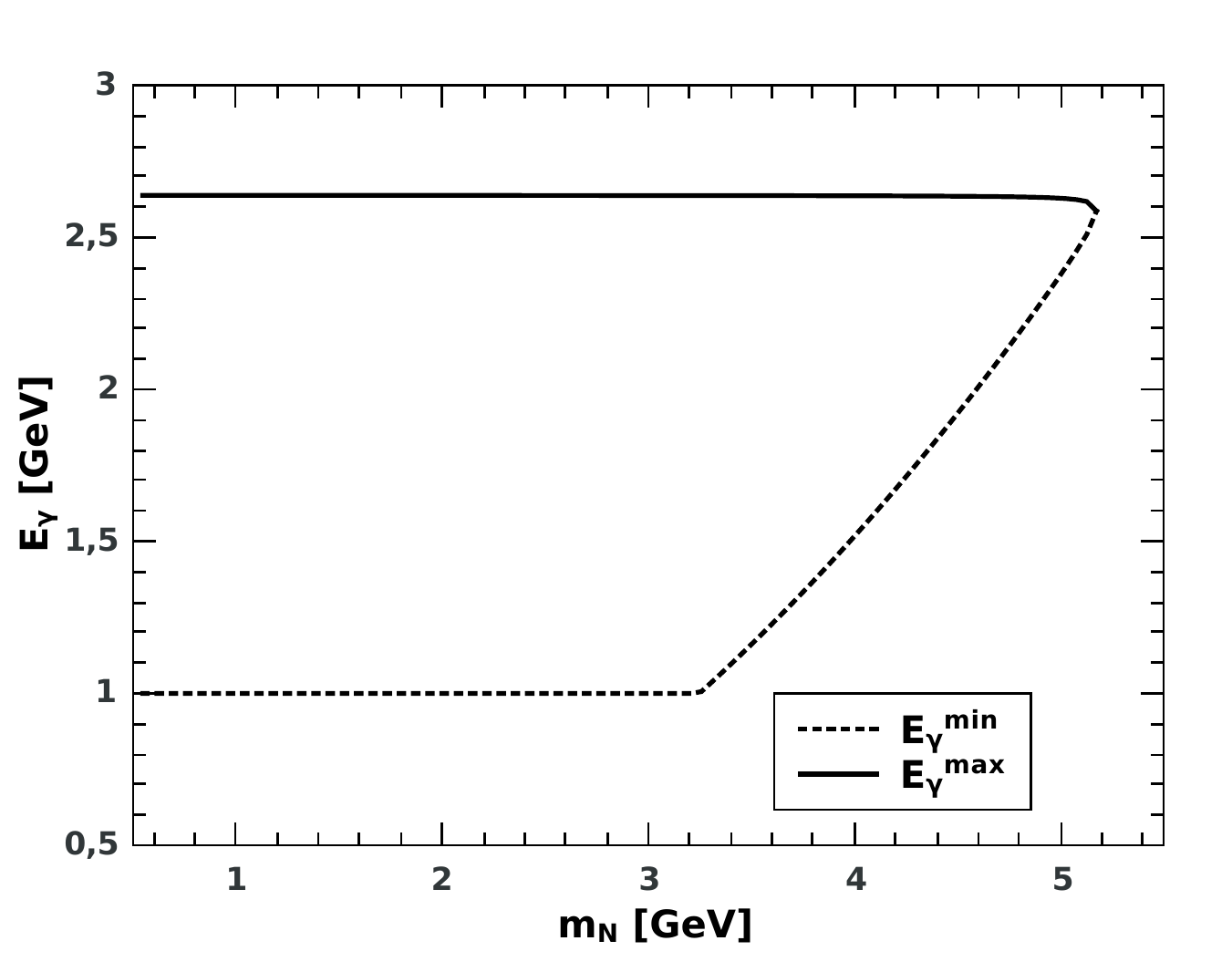}
 \caption{\label{fig:region} Integration limits for $d\Gamma^{(B \rightarrow \mu \nu \gamma)}/d E_{\gamma}$ as a function of $m_N$.}
 \end{figure*}

Now we calculate the contribution of the Majorana-mediated $B$ decay in the effective Lagrangian formalism we want to probe. As we discussed in Sect. \ref{sec:NmedBdec} we consider the process with an intermediate on-shell Majorana neutrino in the Narrow Width Approximation. This process is shown in Fig. \ref{fig:Bmunugamma}. Under these conditions the phase space needs to be organized in order to apply the approximation
\begin{eqnarray}\label{eq:B_N_munug_ps}
d\Gamma^{(B\rightarrow \mu \nu \gamma)}=&\overbrace{\frac1{2m_B}\idotsint \vert \mathcal{M}_{B\to \mu N} \vert^2 (2\pi)^4 \delta^{(4)}(q-p_N-p_1)\delta(p_1^2-m_{\mu}^2)\delta(p_N^2-m_N^2)
\frac{d^4p_1}{(2\pi)^3}\frac{d^4p_N}{(2\pi)^3} }^\text{$\Gamma^{(B\rightarrow \mu N)}$}
\nonumber \\
&\frac1{\Gamma_N}\; \underbrace{\frac1{2 m_N} \vert \mathcal{M}_{N \to \nu \gamma} \vert^2 (2 \pi)^4 \delta^{(4)}(p_N-p_2-k)\delta(p_2^2) \delta(k^2)\frac{d^4p_2}{(2\pi)^3}\frac{d^4k}{(2\pi)^3}}_\text{$d\Gamma^{(N\rightarrow \nu \gamma)}$}.
\end{eqnarray}
Here $\mathcal{M}_{B\to \mu N}$ is the amplitude presented in \eqref{eq:M_BmuN2} and $\mathcal{M}_{N \to \nu \gamma}$ is the amplitude of the radiative $N \rightarrow \nu \gamma$ decay allowed by the one-loop generated operators \eqref{eq:Ope-1-loop} in the Lagrangian \eqref{eq:leff_1loop_L}, again corresponding to the second fermion family $i=2$:   

\begin{eqnarray}\label{eq:M_Nnugamma}
\vert \mathcal{M}_{N \to \nu \gamma}\vert^2= \frac{4 v^2}{\Lambda^4}  m_N^4  (\alpha_{NB}^{(2)} c_W + \alpha^{(2)}_{NW} s_W)^2. 
\end{eqnarray}
Thus, multiplying and dividing \eqref{eq:B_N_munug_ps} by the partial width $\Gamma^{(N\rightarrow \nu \gamma)}$ we have
\begin{eqnarray} \label{eq:dgamma}
d\Gamma^{(B\rightarrow \mu \nu \gamma)}=\Gamma^{(B\rightarrow \mu N)} Br(N\rightarrow \nu \gamma) \frac{d\Gamma^{(N \rightarrow \nu\gamma)}}{\Gamma^{(N \rightarrow \nu\gamma)}}
\end{eqnarray}
where Br($N\rightarrow \nu\gamma$) is the branching ratio in \eqref{eq:Btomunug_decay}.
Partially integrating the phase space, the last factor in \eqref{eq:dgamma} can be written as
\begin{eqnarray}
\frac1{\Gamma^{(N \rightarrow \nu\gamma)}}\frac{d\Gamma^{(N \rightarrow \nu\gamma)}}{dx ~d\cos\theta}=\frac12 \delta(x-1/2)
\end{eqnarray} 
where $x=k^0/m_N$, with $k^0$ the energy of the photon in the Majorana $N$ rest frame. The distribution in the $B$ meson rest frame is obtained by a Lorentz transformation. Here, as in \eqref{eq:Delta_Br_SM} $E_{\gamma}$ is the photon energy in the $B$ rest frame, so
\begin{eqnarray} \label{eq:Lorentz}
E_{\gamma}=k^0 \gamma_N (1+\beta_N \cos\theta) \;\;\text{with} \;\; \gamma_N=\sqrt{1-\beta_N^2} \;\; \text{and} \;\; \beta_N=\sqrt{1-\frac{m_N^2}{E_N^2}}.
\end{eqnarray}
Calling $z=E_{\gamma}/E_N$, where $E_N$ is the Majorana neutrino $N$ energy in the $B$ rest frame we have $z=x(1+\beta_N \cos\theta)$.
We use \eqref{eq:Lorentz} in order to transform the distribution
\begin{eqnarray}
\frac1{\Gamma^{(N \rightarrow \nu\gamma)}}\frac{d\Gamma^{(N \rightarrow \nu\gamma)}}{dz\,dx\,d\cos\theta}=\frac12 \delta(x-1/2)\delta(z-x(1+\beta_N \cos\theta)). 
\end{eqnarray}
Thus, for $-1 < \cos\theta < 1$ we have $\frac12 (1-\beta_N) < z < \frac12 (1+\beta_N)$.
Integrating in $x$ and $\cos\theta$ we have
\begin{eqnarray} \label{eq:dGdE}
&&\frac{d\Gamma^{(B \rightarrow \mu\nu\gamma)}}{dE_{\gamma}}=\Gamma^{(B\rightarrow \mu N)} Br(N\rightarrow \nu \gamma) \frac{1}{E_N \beta_N}
\nonumber \\
\nonumber \\
&&\text{with} \;\; \frac{E_N}2 (1-\beta_N)<E_{\gamma}<\frac{E_N}2 (1+\beta_N).
\end{eqnarray}
In order to obtain the partial branching fraction for $B\rightarrow \mu \nu \gamma$ with photon energy $E_{\gamma}>E_{cut}$ we integrate \eqref{eq:dGdE}
\begin{eqnarray}\label{eq:DeltaBr_eff_}
\Delta Br^{eff}=\int_{E_{\gamma}^{min}}^{E_{\gamma}^{max}} dE_{\gamma} \frac{d\Gamma^{(B\rightarrow \mu \nu \gamma )}}{dE_{\gamma}}=
\Gamma^{(B\rightarrow \mu N )} Br(N\rightarrow \nu\gamma) \frac{(E_{\gamma}^{max}-E_{\gamma}^{min})}{E_N \beta_N}
\end{eqnarray}
where $E_{\gamma}^{min}=\text{max}\left[E_{cut},\frac{E_N}2 (1-\beta_N)\right]$ and $E_{\gamma}^{max}=\frac{E_N}2 (1+\beta_N)$. The integration region is shown in Fig. \ref{fig:region} for $E_{cut}=1$ GeV.


\bibliographystyle{bibstyle.bst}
\bibliography{Bib_N_6_2019}

\end{document}